\documentclass{article}
\usepackage{amssymb}
\usepackage{graphicx}
\usepackage{amsmath}

\input{tcilatex}

\begin{document}

\begin{center}
\bigskip 

{\LARGE Black Holes, Information Loss, and Hidden Variables}

\bigskip

\bigskip\bigskip

\bigskip Antony Valentini\footnote{%
email: avalentini@perimeterinstitute.ca}

\bigskip

\bigskip

\textit{Perimeter Institute for Theoretical Physics, 35 King Street North,
Waterloo, Ontario N2J 2W9, Canada.}

\bigskip
\end{center}

\bigskip

\bigskip

\bigskip

We consider black-hole evaporation from a hidden-variables perspective. It
is suggested that Hawking information loss, associated with the transition
from a pure to a mixed quantum state, is compensated for by the creation of
deviations from Born-rule probabilities outside the event horizon. The
resulting states have non-standard or `nonequilibrium' distributions of
hidden variables, with a specific observable signature -- a breakdown of the
sinusoidal modulation of quantum probabilities for two-state systems.
Outgoing Hawking radiation is predicted to contain statistical anomalies
outside the domain of the quantum formalism. Further, it is argued that even
for a macroscopic black hole, if one half of an entangled EPR-pair should
fall behind the event horizon, the other half will develop similar
statistical anomalies. We propose a simple rule, whereby the relative
entropy of the nonequilibrum (hidden-variable) distribution generated
outside the horizon balances the increase in von Neumann entropy associated
with the pure-to-mixed transition. It is argued that there are relationships
between hidden-variable and von Neumann entropies even in non-gravitational
physics. We consider the possibility of observing anomalous polarisation
probabilities, in the radiation from primordial black holes, and in the
atomic cascade emission of entangled photon pairs from black-hole accretion
discs.

\bigskip

\bigskip

\bigskip

\bigskip

\bigskip

\bigskip

\bigskip

\bigskip

\bigskip

\bigskip

\bigskip

\bigskip

\bigskip

\bigskip

\bigskip

\bigskip

\bigskip

\bigskip

\bigskip

\bigskip

\bigskip

\bigskip

\bigskip

\bigskip

\bigskip

\bigskip

\bigskip

\bigskip

\bigskip

\bigskip

\bigskip

\bigskip

\bigskip

\bigskip

\bigskip

\bigskip

\bigskip

\bigskip

\bigskip

\bigskip

\bigskip

\bigskip

\bigskip

\bigskip

\bigskip

\bigskip

\bigskip

\bigskip

\bigskip

\bigskip

\bigskip

\bigskip

\bigskip

\bigskip

\bigskip

\bigskip

\bigskip

\section{Introduction}

In classical physics, when a star undergoes gravitational collapse to form a
black hole, the final spacetime geometry in the exterior region depends only
on the total mass, charge, and angular momentum. Other details of the
collapsing matter are completely erased. Once the material falls behind the
event horizon, information concerning its detailed structure is screened off
from the exterior.

According to quantum field theory on the resulting classical spacetime
background, the black hole radiates like a body at temperature $T=1/8\pi M$
(in the non-rotating case) \cite{Hawk1}. In conventional units%
\begin{equation}
T=\hbar c^{3}/8\pi GkM\approx (10^{-7}\ \mathrm{K})(M_{\odot }/M)
\label{temp}
\end{equation}%
(where $M_{\odot }$ is the mass of the Sun). The outgoing Hawking radiation
modes are entangled with ingoing modes crossing the horizon. Upon tracing
over the latter degrees of freedom, the field in the asymptotically flat
region is found to be in a mixed state $\hat{\rho}_{\mathrm{ext}}$
corresponding to thermal radiation at the temperature (\ref{temp}).
Heuristically, the process may be visualised in terms of the creation of
entangled pairs of particles near the horizon, with one member of each pair
falling behind the horizon and the other escaping to infinity. Like the
classical exterior geometry, this thermal emission is independent of the
details (apart from the total mass) of what fell behind the horizon.

The emitted radiation decreases the mass of the hole. For slow quasistatic
changes, $T\propto 1/M$ and the rate of energy emission is $\propto 1/M^{2}$%
. The black hole evaporates on a timescale \cite{BHP}%
\begin{equation}
t_{\mathrm{bh}}\sim G^{2}M^{3}/\hbar c^{4}\sim (10^{64}\ \mathrm{yr}%
)(M/M_{\odot })^{3}  \label{tscale}
\end{equation}

Hawking has argued that the formation and evaporation of black holes allows
a closed system to evolve from a pure to a mixed quantum state, in violation
of the usual rules of quantum theory \cite{Hawk2}. Specifically, if matter
in an initial pure state $|\Psi \rangle $ undergoes gravitational collapse,
then once the black hole has evaporated the final mixed state $\hat{\rho}_{%
\mathrm{ext}}$ becomes the state of the whole system. According to this
argument, the formation and evaporation of the black hole results in a
pure-to-mixed evolution, represented by a non-unitary map $|\Psi \rangle
\rightarrow \hat{\rho}_{\mathrm{ext}}$. Further, because $\hat{\rho}_{%
\mathrm{ext}}$ describes thermal radiation that depends on the initial mass
of the hole but is independent of the details of $|\Psi \rangle $, it
follows that given the final state $\hat{\rho}_{\mathrm{ext}}$ it is
impossible to retrodict the initial state $|\Psi \rangle $ -- a situation
that is commonly referred to as `information loss', since the details of
what originally fell behind the horizon seem to have been erased.

Hawking's argument for information loss is controversial, and a number of
well-known proposals have been made to avoid the conclusion of pure-to-mixed
evolution. (For reviews see, for example, refs. \cite%
{Pres93,Pag94,Gid95,Str96,Bel99}.) It is sometimes suggested that
evaporation stops as the hole approaches the Planck mass, leaving a remnant
such that the total state is still pure; however, since the initial mass can
be arbitrarily large, such low-mass remnants would have to have an
arbitrarily large number of internal states, arguably leading to unbounded
production rates for remnants in other (soft) processes \cite{Gid95}.
Alternatively, it has been suggested that the emitted radiation contains
extra correlations over time such that the final radiated state is actually
pure; but to produce such correlations seems to require nonlocal
interactions operating across the horizon \cite{Wald94}. Another possibility
is that during evaporation a new universe is formed, causally disconnected
from our own, such that the joint state of the two universes is still pure;
this scenario, too, has its difficulties \cite{Pres93,Pag94,Str96,Bel99,PS94}%
.

The desire to avoid pure-to-mixed transitions has been a strong motivation
for the holographic hypothesis, according to which the fundamental degrees
of freedom are defined on a boundary of conventional spacetime \cite{Holog}.
Holography, and the closely-related idea of black-hole complementarity, make
it seem possible that the information that apparently disappears behind the
horizon is actually encoded on the horizon (from where it can be transferred
to the outgoing radiation). A possible realisation of this is through
AdS/CFT duality \cite{AdS}, which relates a string theory in ten dimensions
to a gauge field theory in four dimensions (the local gauge-theory
observables being mapped to boundary conditions on the string spacetime).
According to this framework, the presumed unitarity of the dual gauge
description guarantees that the formation and evaporation of black holes
will be strictly unitary, with appropriate correlations in the outgoing
Hawking radiation; though it remains to be shown explicitly -- in terms of
the gravitational variables -- where the (semiclassical) argument for a
pure-to-mixed transition breaks down \cite{Das00,Pol}.

Other authors conclude that the argument for information loss may signal a
genuine failure of quantum theory \cite{Pres93}. Hawking \cite{Hawk2} has
proposed that quantum theory should be generalised to include pure-to-mixed
evolution. It was suggested that such evolution would imply
currently-observable violations of either locality or energy-momentum
conservation \cite{Ban84}; but in fact this is not necessarily the case \cite%
{Un95}. Penrose \cite{Pen} has argued (on the basis of thermal fluctuations
for black holes in equilibrium with radiation in a large container) that
information loss should be balanced by a gravitationally-induced collapse of
the state vector, which might be observable for massive bodies \cite{Mar03}.
It has been suggested by 't Hooft \cite{tH99} that information loss in black
holes might be clarified in a deterministic but dissipative local
hidden-variables theory (that is, in a local and deterministic completion of
quantum theory with effective irreversibility at the hidden-variable level);
though it is unclear how a local hidden-variables theory can be reconciled
with the observed violations of Bell's inequalities \cite{tH02}. Recently,
it has been suggested that entangled states might in principle be used to
test for nonlinear evolution \cite{Yurt03} and for (`superquantum') cloning
or deleting \cite{Sen04} behind the horizon, by allowing part of the state
to fall into the hole and monitoring the exterior.

It is sometimes suggested that Hawking's argument points to the need for
some form of nonlocal information flow from behind the horizon, throughout
the evaporation process (even when the black hole is macroscopic) \cite%
{Dan93}. Holography and black-hole complementarity are often regarded as
effective instantiations of this, whereby locality breaks down not merely
microscopically but over macroscopic distances comparable to the size of the
hole. For example, a failure of commutativity over spacelike separations
(for observables inside and outside the horizon) might arise from strings
stretched between spacelike-separated points \cite{Low95,Pol95}.\footnote{%
In this context see also, for example, refs. \cite{Mat04,Kap04,Gid04} and
references therein.} According to holography, the fundamental degrees of
freedom are nonlocal, and locality in ordinary spacetime emerges only in
some approximation; but again, the precise nonlocal mechanism that leads to
purity of the outgoing state is not easy to discern in the semiclassical
regime, because of the difficulty in translating the dual picture into the
higher-dimensional picture \cite{Das00,Pol}.

In this paper, we shall assume that the pure-to-mixed transition $|\Psi
\rangle \rightarrow \hat{\rho}_{\mathrm{ext}}$ for quantum states really
does occur during the formation and evaporation of black holes.
Nevertheless, we shall argue that information need not be lost (in the sense
that retrodiction might still be possible) if one allows deviations from the
standard Born rule for quantum probabilities. For entangled states, such
deviations may depend nonlocally on processes in the interior, so that
information does indeed escape from behind the horizon.

To motivate our proposal, consider an analogy with classical statistical
thermodynamics. When an ideal gas reaches thermal equilibrium, the resulting
macroscopic state contains no memory of how the state was formed; the
approach to statistical equilibrium has the remarkable effect of erasing
information about the past, at least at the macroscopic level. Furthermore,
once the gas has reached thermal equilibrium, its macroscopic state may be
specified by a very small number of parameters, such as volume and
temperature. And all statistical information about the gas may be obtained
from the partition function $Z=\sum_{E}e^{-E/kT}$, again regardless of how
the equilibrium state was prepared.

Now, in classical physics this `thermal information loss' would be regarded
as an artifact of averaging over microscopic degrees of freedom. At any
finite time, the precise positions and velocities of the gas molecules
contain more detailed information from which the details of the past
preparation may in principle be recovered (assuming a deterministic and
time-reversible dynamics).

With this in mind one might consider that, in the case of black holes,
information about what fell behind the horizon is actually stored in extra
degrees of freedom that are usually disregarded (or averaged over). Our
proposal is that these extra degrees of freedom are nonlocal hidden
variables -- parameters of a nonlocal deterministic theory which quantum
theory averages over (much as classical statistical thermodynamics averages
over the deterministic dynamics of gas molecules).

Deviations from the Born rule are a natural possibility in deterministic
hidden-variables theories. For example, in the pilot-wave theory of de
Broglie and Bohm [29--38], the quantum state of an individual system is
supplemented by a (hidden) deterministic trajectory in configuration space,
with velocity given by the gradient of the phase of the wave function. For
an ensemble of systems with wave function $\psi $, it is usually assumed
that the configurations are distributed according to the Born rule $\rho
=|\psi |^{2}$. This assumption (made at some initial time) guarantees
empirical agreement with the statistical predictions of quantum theory. But
there is no reason -- within the theory -- why one could not consider more
general, `nonequilibrium' distributions $\rho \neq |\psi |^{2}$ (just as in
classical physics one may consider ensembles that depart from thermal
equilibrium). These lead to predictions that deviate from standard Born-rule
probabilities for outcomes of quantum measurements; and, it is found that
the marginal statistics at one wing of an entangled state generally depend
nonlocally on what happens at the other wing \cite{PLA2}.

More generally, deterministic hidden-variables theories contain extra
degrees of freedom $\lambda $ that are averaged over some `quantum
equilibrium' distribution $\rho _{\mathrm{eq}}(\lambda )$ to yield the
statistical predictions of quantum theory \cite{Bell66}. Under reasonable
assumptions, Bell's theorem requires that all such theories be nonlocal \cite%
{Bell64}. For generic `quantum nonequilibrium' distributions $\rho (\lambda
)\neq \rho _{\mathrm{eq}}(\lambda )$, the outcomes of quantum measurements
violate the Born rule \cite{Sig}, and the underlying nonlocality is visible
at the statistical level \cite{PLAc,Cracow}.

At present there is of course no evidence for deviations from the Born rule,
and in discussions of hidden-variables theories attention is usually
restricted to the quantum equilibrium state. This state appears to be
stable, at least in non-gravitational physics.

We suggest that quantum equilibrium is disturbed in the interior of black
holes (perhaps near the singularity). For entangled quantum states that
straddle the horizon, deviations from equilibrium in the interior may (as we
shall see) be transmitted to the exterior region, resulting in observable
deviations from standard quantum probabilities outside the black hole. This
provides a possible mechanism by which information from behind the horizon
could indeed reach the exterior by nonlocal effects, though in a way that is
quite different from previous proposals.

We shall argue, then, that nonlocal hidden variables provide additional
degrees of freedom in which information may be stored about what fell behind
the event horizon of an evaporating black hole, and that via entanglement
this information may leak out to the exterior region. In effect, the hidden
variables provide an additional entropy reservoir that is usually filled (in
equilibrium), and becomes `unfrozen' in nonequilibrium.

It will be suggested that the transition $|\Psi \rangle \rightarrow \hat{\rho%
}_{\mathrm{ext}}$ from a pure to a mixed quantum state is accompanied by the
creation of anomalous nonequilibrium distributions $\rho (\lambda )\neq \rho
_{\mathrm{eq}}(\lambda )$ of hidden variables outside the horizon. Such
distributions carry a specific signature, in the form of a breakdown of the
sinusoidal modulation of quantum probabilities for two-state systems \cite%
{Sig}. A simple rule will be proposed, whereby the relative entropy of the
nonequilibrium distribution generated outside the horizon balances the
increase in von Neumann entropy associated with the transition $|\Psi
\rangle \rightarrow \hat{\rho}_{\mathrm{ext}}$, yielding quantitative
predictions that could (at least in principle) be tested experimentally.

According to our proposal, outgoing Hawking radiation will contain
statistical anomalies outside the domain of the quantum formalism --
anomalies corresponding to non-standard distributions of hidden variables.
In particular, photons will have anomalous polarisation probabilities,
deviating from the $\cos ^{2}\Theta $ modulation of transmission through a
pair of polarisers set at a relative angle $\Theta $. Further, we shall
argue that similar anomalies could be created by allowing one half of an
entangled EPR-pair to fall behind the event horizon of a macroscopic black
hole. As we shall see, it is not impossible that these effects could be
observed in radiation from primordial black holes, and in photons from
atomic cascade emissions in black-hole accretion discs.

The prediction of thermal radiation from black holes has long been thought
to point to a deep connection between quantum theory, gravitation, and
statistical physics. And Hawking's argument, that the formation and
evaporation of black holes can induce a transition from a pure to a mixed
quantum state, still stands as a challenge (albeit a controversial one) to
the basic principles of quantum theory. According to the reasoning given
below, the connection with gravitation involves not just ordinary
statistical physics, but the statistical physics of nonlocal hidden
variables.

In section 2, we review the notion of statistical equilibrium in classical
and pilot-wave dynamics, and in deterministic hidden-variables theories
generally. It is shown that quantum equilibrium has special properties, such
as locality, which arise from an effective erasure of underlying degrees of
freedom. We also summarise what is currently known about gravitation in the
context of hidden-variables theories. It is emphasised that while
hidden-variables theories with a stable equilibrium state may easily be
constructed on a globally-hyperbolic spacetime (an example is given), it is
not known whether this can be done in the non-globally-hyperbolic case.

In section 3, we consider a thought experiment with black holes and
entangled states. We make the hypothesis that for a quantum state entangled
across the horizon, the external part of the state evolves away from quantum
equilibrium, in accordance with a simple rule. A model is provided, showing
how an entangled state can provide a channel for the nonlocal propagation of
nonequilibrium across the horizon. It is noted that nonequilibrium may be
conveniently detected in the form of anomalous polarisation probabilities
for photons, and that the expected degree of nonequilibrium is related in a
simple way to the maximal violation of Bell's inequality associated with the
entangled state.

In section 4, the hypothesis of section 3 is extended to the entangled field
modes of ingoing and outgoing Hawking radiation, resulting in a constraint
on the quantum nonequilibrium distribution for particles emitted by black
holes.

In section 5, we discuss how the seemingly disparate concepts of
hidden-variable and von Neumann entropies are related even in
non-gravitational physics. In particular, systems in quantum nonequilibrium
(with non-zero hidden-variable entropy) may be used to separate
non-orthogonal quantum states for ordinary systems, resulting in an
anomalous evolution of the von Neumann entropy.

In section 6, we consider the possibility of observing the proposed
processes experimentally, in the Hawking radiation from primordial black
holes (perhaps left over from the early universe), and in entangled photons
from black-hole accretion discs. The latter possibility in particular is
considered in some detail. We discuss current observations of iron emission
lines in the vicinity of the event horizons of macroscopic black holes, and
argue that the identification of an atomic cascade generating entangled
photon pairs close to a horizon would enable the experiment to be performed,
at least in principle.

\section{Thermal and Quantum Equilibrium}

In this section, we first review the notion of statistical equilibrium in
deterministic theories -- in classical and pilot-wave dynamics, and in
general (deterministic) hidden-variables theories. We note some special
properties of equilibrium. In particular, we show how equilibrium erases
information about underlying degrees of freedom. Finally, we sketch what is
currently known about the role of gravitation in hidden-variables theories.

\subsection{Classical and Pilot-Wave Dynamics}

In classical dynamics, the phase-space trajectory $\left( q(t),p(t)\right) $
of an individual system is determined by Hamilton's equations $\dot{q}%
=\partial H/\partial p$ and $\dot{p}=-\partial H/\partial q$, given the
initial conditions $\left( q_{0},p_{0}\right) $. For an ensemble with the
same Hamiltonian, the velocity field $\dot{X}\equiv \left( \dot{x},\dot{p}%
\right) $ determines the evolution of any distribution $\rho (q,p,t)$ via
the continuity equation (with $\nabla \equiv \left( \partial /\partial
q,\partial /\partial p\right) $)%
\begin{equation}
\frac{\partial \rho }{\partial t}+\nabla \cdot (\rho \dot{X})=0
\label{Contcl}
\end{equation}%
(more usually written as $\partial \rho /\partial t+\left\{ \rho ,H\right\}
=0$). Because $\nabla \cdot \dot{X}=0$, a uniform initial distribution $\rho
(q,p,0)=\mathrm{const}.$ (on the energy surface) remains uniform, $\rho
(q,p,t)=\mathrm{const}.$. This steady state is just thermal equilibrium. For
an arbitrary initial state $\rho (q,p,0)$, the evolution $\rho (q,p,t)$ is
obtained -- in principle -- from integration of (\ref{Contcl}).

The deviation of $\rho $ from thermal equilibrium may be quantified in terms
of the classical \textit{H}-function%
\begin{equation}
H_{\mathrm{class}}=\int \int dqdp\ \rho \ln \rho  \label{Hcl}
\end{equation}%
Under Hamiltonian evolution, Liouville's theorem states that $d\rho /dt=0$
along trajectories, so that the exact $H_{\mathrm{class}}$ is constant in
time, $dH_{\mathrm{class}}/dt=0$. But in appropriate circumstances, the
coarse-grained \textit{H}-function%
\begin{equation}
\bar{H}_{\mathrm{class}}=\int \int dxdp\;\bar{\rho}\ln \bar{\rho}
\label{Hbarcl}
\end{equation}%
does decrease, corresponding to thermal relaxation on a coarse-grained
level. Assuming there is no initial fine-grained microstructure in $\rho $
at $t=0$, we have the classical coarse-graining \textit{H}-theorem \cite%
{Tol,Dav} $\bar{H}_{\mathrm{class}}(t)\leq \bar{H}_{\mathrm{class}}(0)$,
where $\bar{H}_{\mathrm{class}}$ is minimised by $\bar{\rho}=\mathrm{const}.$%
. The \textit{H}-theorem formalises the idea of Gibbs \cite{Gibbs} -- that
an initial non-uniform distribution tends to develop fine-grained structure,
becoming more uniform on a coarse-grained level.

In pilot-wave theory, the dynamics takes place in configuration space. The
trajectory $q(t)$ of an individual system is determined by the wave function 
$\psi (q,t)$ via the de Broglie guidance equation%
\begin{equation}
\frac{dq}{dt}=\frac{j}{\left\vert \psi \right\vert ^{2}}  \label{deB}
\end{equation}%
where $\psi $ obeys the usual Schr\"{o}dinger equation%
\begin{equation}
i\frac{\partial \psi }{\partial t}=\hat{H}\psi  \label{Sch}
\end{equation}%
in configuration space (units $\hbar =1$), and where $j=j\left[ \psi \right]
=j(q,t)$ is the conserved current derived from (\ref{Sch}), satisfying%
\begin{equation}
\frac{\partial \left\vert \psi \right\vert ^{2}}{\partial t}+\nabla \cdot j=0
\label{Contj}
\end{equation}%
(where here $\nabla \equiv \partial /\partial q$). For example, for a system
of $n$ nonrelativistic particles with positions $\mathbf{x}_{i}(t)$ and
masses $m_{i}$, we have $q=(\mathbf{x}_{1},\mathbf{x}_{2},....,\mathbf{x}%
_{n})$ and (\ref{deB}) generally takes the form%
\begin{equation}
\frac{d\mathbf{x}_{i}}{dt}=\frac{1}{m_{i}}\func{Im}\frac{\mathbf{\nabla }%
_{i}\psi }{\psi }=\frac{\mathbf{\nabla }_{i}S}{m_{i}}  \label{deBpart}
\end{equation}%
where $\psi =\left\vert \psi \right\vert e^{iS}$.

Mathematically, $j/\left\vert \psi \right\vert ^{2}$ is the ratio of the
quantum probability current to the quantum probability density. Physically,
however, $\psi $ (and hence $j[\psi ]$) is here regarded as an objective
physical field (in configuration space) guiding the motion of an individual
system.

The equations (\ref{deB}) and (\ref{Sch}) define a deterministic dynamics
for individual systems. Given the initial field (or wave function) $\psi
(q,0)$, (\ref{Sch}) determines $\psi (q,t)$ at all times. And given the
initial configuration $q(0)$, (\ref{deB}) then determines the trajectory $%
q(t)$ at all times. Note that $\psi $ has no \textit{a priori} connection
with probabilities: it is a physical field on configuration space, driving
the dynamics of an individual system.

For an ensemble of independent systems, each with the same wave function $%
\psi (q,t)$, we may define a distribution $\rho (q,t)$ of actual
configurations $q$ at time $t$. In principle, the ensemble distribution $%
\rho (q,t)$ need have no relation to $|\psi (q,t)|^{2}$.

The guidance equation (\ref{deB}) defines a velocity field $\dot{q}$, which
determines the evolution of any distribution $\rho (q,t)$ via the continuity
equation

\begin{equation}
\frac{\partial \rho }{\partial t}+\nabla \cdot \left( \rho \dot{q}\right) =0
\label{cont}
\end{equation}%
(where again $\nabla \equiv \partial /\partial q$).

Rewriting (\ref{Contj}) as%
\begin{equation}
\frac{\partial \left\vert \psi \right\vert ^{2}}{\partial t}+\nabla \cdot
(\left\vert \psi \right\vert ^{2}\dot{q})=0  \label{Contpsi2}
\end{equation}%
it follows that an initial distribution $\rho (q,0)=\left\vert \psi
(q,0)\right\vert ^{2}$ evolves into $\rho (q,t)=\left\vert \psi
(q,t)\right\vert ^{2}$. This is the state of quantum equilibrium, analogous
to thermal equilibrium.

Given the velocity field $\dot{q}$, (\ref{cont}) determines the evolution of
an \textit{arbitrary} initial distribution $\rho (q,0)\neq \left\vert \psi
(q,0)\right\vert ^{2}$.

The deviation of $\rho $ from quantum equilibrium may be quantified in terms
of the hidden-variable \textit{H}-function \cite{AVth,PLA1,AVIsch}%
\begin{equation}
H_{\mathrm{hv}}=\int dq\ \rho \ln (\rho /|\psi |^{2})  \label{Hfn}
\end{equation}%
From (\ref{cont}) and (\ref{Contpsi2}), the ratio $f=\rho /|\psi |^{2}$ is
preserved along trajectories, $df/dt=0$ (the analogue of Liouville's
theorem), so that the exact $H_{\mathrm{hv}}$ is constant in time, $dH_{%
\mathrm{hv}}/dt=0$. However, as in the classical case, in appropriate
circumstances the coarse-grained \textit{H}-function%
\begin{equation}
\bar{H}_{\mathrm{hv}}=\int dq\ \bar{\rho}\ln (\bar{\rho}/\overline{%
\left\vert \psi \right\vert ^{2}})  \label{Hbarhv}
\end{equation}%
does decrease, corresponding to relaxation to quantum equilibrium on a
coarse-grained level. Assuming there is no initial fine-grained
microstructure in $\rho $ and $|\psi |^{2}$ at $t=0$, we have the
coarse-graining \textit{H}-theorem \cite{AVth,PLA1,AVIsch} $\bar{H}_{\mathrm{%
hv}}(t)\leq \bar{H}_{\mathrm{hv}}(0)$, where $\bar{H}_{\mathrm{hv}}\geq 0$
for all $\bar{\rho}$, $\overline{\left\vert \psi \right\vert ^{2}}$ and $%
\bar{H}_{\mathrm{hv}}=0$ if and only if $\bar{\rho}=\overline{\left\vert
\psi \right\vert ^{2}}$ everywhere. This version of the \textit{H}-theorem
formalises the idea that $\rho $ and $\left\vert \psi \right\vert ^{2}$
behave like two `fluids' which are `stirred' by the same velocity field $%
\dot{q}$, so that $\rho $ and $\left\vert \psi \right\vert ^{2}$ tend to
become indistinguishable on a coarse-grained level.

Note that the `hidden-variable entropy'%
\begin{equation}
S_{\mathrm{hv}}=-\int dq\ \rho \ln (\rho /\left\vert \psi \right\vert ^{2})
\label{Enthv}
\end{equation}%
is just the relative entropy of $\rho $ with respect to $\left\vert \psi
\right\vert ^{2}$, and is a natural measure of the difference between $\rho $
and $\left\vert \psi \right\vert ^{2}$.

Significant relaxation $\rho \rightarrow \left\vert \psi \right\vert ^{2}$
(on a coarse-grained level) occurs only if the velocity field varies rapidly
over the coarse-graining cells. For very simple systems, such as a single
particle in an energy eigenstate (for which the velocity field (\ref{deB})
vanishes), there is no relaxation at all -- just as there is none for an
ensemble of classical particles bouncing back and forth perpendicular to the
walls of a box. For systems whose wave functions are a superposition of many
energy states, the velocity field varies rapidly and numerical simulations
confirm the expected relaxation, on timescales that agree with the estimate
obtained from time-derivatives of $\bar{H}_{\mathrm{hv}}(t)$ near $t=0$ \cite%
{AVIsch,AVHW}.

A statistical mixture of wave functions $\psi _{\alpha }(q,t)$, weighted by
probabilities $p_{\alpha }$, corresponds to a mixed quantum state with
density operator%
\begin{equation*}
\hat{\rho}=\sum_{\alpha }p_{\alpha }|\psi _{\alpha }\rangle \langle \psi
_{\alpha }|
\end{equation*}%
For each pure subensemble with wave function $\psi _{\alpha }(q,t)$, one may
define a distribution $\rho _{\alpha }(q,t)$ (which need not equal $|\psi
_{\alpha }(q,t)|^{2}$) and an \textit{H}-function%
\begin{equation*}
H_{\mathrm{hv}}^{\alpha }=\int dq\;\rho _{\alpha }\ln (\rho _{\alpha
}/\left\vert \psi _{\alpha }\right\vert ^{2})
\end{equation*}%
satisfying the above theorem. For the whole ensemble, with distribution%
\begin{equation*}
\rho (q,t)=\sum_{\alpha }p_{\alpha }\rho _{\alpha }(q,t)
\end{equation*}%
one may define a mean \textit{H}-function%
\begin{equation}
H_{\mathrm{hv}}=\sum_{\alpha }p_{\alpha }H_{\mathrm{hv}}^{\alpha }
\label{MeanHhv}
\end{equation}%
and again the coarse-grained $H_{\mathrm{hv}}$ will satisfy the theorem (for
a closed system with constant $p_{\alpha }$). In equilibrium, $H_{\mathrm{hv}%
}=0$ implies $H_{\mathrm{hv}}^{\alpha }=0$ and $\rho _{\alpha }=|\psi
_{\alpha }|^{2}$ (for all $\alpha $), the distribution for the whole
ensemble then being equal to $\rho (q,t)=\langle q|\hat{\rho}(t)|q\rangle $.
Thus, for mixed states the hidden-variable entropy is%
\begin{equation}
S_{\mathrm{hv}}=-\sum_{\alpha }p_{\alpha }\int dq\;\rho _{\alpha }\ln (\rho
_{\alpha }/\left\vert \psi _{\alpha }\right\vert ^{2})  \label{Enthv2}
\end{equation}%
Note that in pilot-wave theory, mixed quantum states are interpreted as
statistical mixtures of physically real pilot waves, which correspond to a
preferred decomposition of $\hat{\rho}$.

The above statistical mechanics of hidden variables is conceptually similar
to its classical counterpart. In the time-reversal invariant dynamics
defined by (\ref{deB}) and (\ref{Sch}), for every initial state that evolves
towards equilibrium one can construct a time-reversed `initial' state that
evolves away from equilibrium. Such reversed initial states will, however,
contain fine-grained microstructure. Thus, relaxation to equilibrium
requires an assumption about initial conditions, and any such assumption is
arguably related to questions of cosmology \cite%
{Dav,AVIsch,AVHW,Sklar,Hall,Savitt}.

It has been suggested that the universe began in a state of quantum
nonequilibrium, the relaxation $\rho \rightarrow \langle q|\hat{\rho}%
|q\rangle $ taking place during the violence of the big bang, and that
remnants of early nonequilibrium might be found in relic particles that
decoupled at sufficiently early times \cite{AVth,AVbook,PLA1,AVIsch,AV96,NS}%
. An alternative view \cite{DGZ92} takes $|\psi |^{2}$ to be the natural
measure of probability or `typicality' for initial configurations of the
whole universe (with $\psi $ the universal wave function), resulting in
quantum probabilities for all subsystems at all times.

For the purposes of the present paper, we may leave aside the question of
initial conditions. The key point is that one can in principle consider
nonequilibrium distributions -- within the framework of de Broglie-Bohm
theory, and indeed in any deterministic hidden-variables theory (just as it
is possible to consider thermal nonequilibrium in classical physics). And
the natural measure of quantum nonequilibrium is the hidden-variable entropy
(\ref{Enthv}), or (\ref{Enthv2}) for mixed states.

As a simple example of quantum nonequilibrium, consider an ensemble of free
nonrelativistic particles represented by the quantum momentum state $%
\left\vert \mathbf{p}\right\rangle $, where the particles have energy $%
E=p^{2}/2m$ and are confined to a large normalisation volume $V$. The state $%
\left\vert \mathbf{p}\right\rangle $ has wave function $\psi (\mathbf{x}%
,t)=e^{i\mathbf{p}\cdot \mathbf{x}}e^{-iEt}/\sqrt{V}$. Should such an
ensemble be prepared experimentally, and should the particle positions then
be measured, quantum theory predicts a uniform distribution $|\psi |^{2}=1/V$
of measured results. In pilot-wave theory there is no reason (in principle)
why the measured particle distribution could not, for example, be confined
to one half of the volume $V$. Similarly, a plane wave $\psi _{\mathrm{inc}}$
incident on a two-slit screen yields the usual interference pattern at the
backstop, provided the incident particles have distribution $\rho _{\mathrm{%
inc}}=|\psi _{\mathrm{inc}}|^{2}$. It is a well-known feature of pilot-wave
theory that incoming particles on one side of the symmetry axis
(perpendicular to the screen) hit the backstop on the same side of the
symmetry axis \cite{Bell,Holl}. Thus if, for example, the incident
distribution $\rho _{\mathrm{inc}}$ has support on only one side of the
symmetry axis, all particles will land on only one half of the backstop, and
only one half of the interference pattern will appear.

These examples make it clear that, at least in principle, pilot-wave theory
with $\rho =|\psi |^{2}$ is a special case of a wider theory (just as
classical physics in thermal equilibrium is a special case of a wider
theory). For nonequilibrium distributions $\rho \neq |\psi |^{2}$ at $t=0$,
the outcomes of subsequent quantum measurements over an ensemble will have a
distribution that departs from quantum predictions (assuming relaxation $%
\rho \rightarrow |\psi |^{2}$ has not occurred by the time the measurements
have taken place).

\subsection{General (Deterministic) Hidden-Variables Theories}

The situation is conceptually the same in any deterministic hidden-variables
theory. Consider, for example, a two-state system in the standard
Bloch-sphere representation. A quantum measurement of the observable $\hat{%
\sigma}\equiv \mathbf{m}\cdot \mathbf{\hat{\sigma}}$ (which might be spin
along an axis $\mathbf{m}$ in space, in units of $\hbar /2$) can yield
outcomes $\sigma =\pm 1$. In a deterministic hidden-variables theory, the
outcome $\sigma $ is determined in advance by the setting $\mathbf{m}$ of
the measuring apparatus together with some unknown parameters $\lambda $
(defined at some initial time $t=0$, say at preparation). In other words,
there is a deterministic mapping%
\begin{equation}
\sigma =\sigma (\mathbf{m},\lambda )  \label{dethv}
\end{equation}%
from the conditions $\mathbf{m},\lambda $ to the outcome $\sigma =\pm 1$.

Over an ensemble, with fixed measurement axis $\mathbf{m}$ and variable $%
\lambda $, quantum theory can hold only if there exists a distribution $\rho
_{\mathrm{eq}}(\lambda )$ of hidden variables such that the mean outcome%
\footnote{%
It is customary to write as if $\lambda $ were a continuous variable. The
integral sign is really a generalised sum, and no assumptions are made about 
$\lambda $.}%
\begin{equation*}
\left\langle \sigma \left( \mathbf{m},\lambda \right) \right\rangle _{%
\mathrm{eq}}\equiv \int d\lambda \ \rho _{\mathrm{eq}}(\lambda )\sigma
\left( \mathbf{m},\lambda \right)
\end{equation*}%
is equal to the quantum mean%
\begin{equation}
\left\langle \mathbf{m}\cdot \mathbf{\hat{\sigma}}\right\rangle =Tr\left( 
\hat{\rho}\mathbf{m}\cdot \mathbf{\hat{\sigma}}\right) =\mathbf{m}\cdot
Tr\left( \hat{\rho}\mathbf{\hat{\sigma}}\right) \equiv \mathbf{m}\cdot 
\mathbf{P}=P\cos \theta  \label{Qm}
\end{equation}%
where $\hat{\rho}$ is the quantum density operator associated with the
preparation, $\mathbf{P}$ is the mean polarisation, and $\theta $ is the
angle (on the Bloch sphere) between $\mathbf{m}$ and $\mathbf{P}$. With just
two possible outcomes $\sigma =\pm 1$, the mean fixes the outcome
probabilities%
\begin{equation}
p_{\mathrm{eq}}^{\pm }(\mathbf{m})=\frac{1}{2}\left( 1\pm P\cos \theta
\right)  \label{Qp}
\end{equation}%
which depend sinusoidally on $\theta $, for any pure or mixed state (as long
as $P\neq 0$). If the ensemble is completely polarised, $P=1$ and $p_{%
\mathrm{eq}}^{+}(\mathbf{m})=\cos ^{2}(\theta /2)$.

Thus, any deterministic hidden-variables theory is specified by two
conceptually distinct components. The first refers to individual systems:
the mapping (\ref{dethv}) determines the outcome $\sigma $, given the
conditions $\mathbf{m},\lambda $. The second refers to an ensemble: the
distribution $\rho _{\mathrm{eq}}(\lambda )$ specifies how the individual
hidden variables $\lambda $ are distributed over the ensemble.\footnote{%
The uncertainty principle -- that is, the unavoidable statistical dispersion
over quantum ensembles -- originates from the dispersion of $\rho _{\mathrm{%
eq}}(\lambda )$.}

Conceptually, this is the same as in pilot-wave theory: (\ref{deB}) and (\ref%
{Sch}) determine the outcome of any quantum experiment, given the `hidden
variables' $q(0)$, $\psi (q,0)$ together with the experimental arrangement
(specified by the external potential in (\ref{Sch})); while the choice $\rho
(q,0)=|\psi (q,0)|^{2}$ for the initial ensemble distribution guarantees
that the outcome probabilities agree with quantum theory.

Clearly, just as one may contemplate non-standard ensemble distributions $%
\rho (q,0)\neq |\psi (q,0)|^{2}$ in pilot-wave theory, so one may equally
contemplate non-standard ensemble distributions $\rho (\lambda )\neq \rho _{%
\mathrm{eq}}(\lambda )$ in any deterministic hidden-variables theory. By
retaining the same mapping (\ref{dethv}) for individual systems, one then
obtains a theory that is wider than quantum theory but includes it as a
special case: quantum probabilities are obtained for the `equilibrium'
distribution $\rho (\lambda )=\rho _{\mathrm{eq}}(\lambda )$, but not for
general `nonequilibrium' distributions $\rho (\lambda )\neq \rho _{\mathrm{eq%
}}(\lambda )$.

Generically, the nonequilibrium ensemble mean will deviate from the quantum
prediction,%
\begin{equation}
\left\langle \sigma \left( \mathbf{m},\lambda \right) \right\rangle \equiv
\int d\lambda \ \rho (\lambda )\sigma \left( \mathbf{m},\lambda \right) \neq
P\cos \theta  \label{NQm}
\end{equation}%
and the outcome probabilities%
\begin{equation}
p^{\pm }(\mathbf{m})=\frac{1}{2}\left( 1\pm \left\langle \sigma \left( 
\mathbf{m},\lambda \right) \right\rangle \right) \neq \frac{1}{2}\left( 1\pm
P\cos \theta \right)  \label{NQp}
\end{equation}%
will \textit{not} be sinusoidal in $\theta $ \cite{Sig}.

The essential point is that in any deterministic theory (including classical
or pilot-wave dynamics) there is a clear conceptual distinction between the
dynamical equations for an individual system and the distribution of initial
conditions over an ensemble. One is free to retain the former while changing
the latter.

In a general hidden-variables theory, with nonequilibrium distributions $%
\rho (\lambda )\neq \rho _{\mathrm{eq}}(\lambda )$, it is difficult to see
how one could construct an argument for relaxation $\rho (\lambda
)\rightarrow \rho _{\mathrm{eq}}(\lambda )$, in the absence of specific
dynamical equations. Nevertheless, the key point is generally true: any
deterministic hidden-variables theory with $\rho (\lambda )=\rho _{\mathrm{eq%
}}(\lambda )$ is a special case of a wider theory with generic
nonequilibrium $\rho (\lambda )\neq \rho _{\mathrm{eq}}(\lambda )$. And the
natural measure of nonequilibrium is the hidden-variable entropy%
\begin{equation}
S_{\mathrm{hv}}=-\int d\lambda \ \rho \ln (\rho /\rho _{\mathrm{eq}})
\label{Enthv3}
\end{equation}%
(the relative entropy of $\rho (\lambda )$ with respect to $\rho _{\mathrm{eq%
}}(\lambda )$).

\subsection{Properties of Equilibrium. Local Statistics, Memory Loss, and
Information Compression}

In quantum theory, for an entangled state of two widely-separated systems,
the marginal statistics of outcomes at each wing do not depend on the
measurement setting at the other distant wing. This locality of statistics
is remarkable from a hidden-variables perspective, given that any reasonable
hidden-variables theory has to be fundamentally nonlocal.

Specifically, consider a pair of two-state systems at points $A$ and $B$ in
space. Quantum measurements of the local observables $\hat{\sigma}_{A}\equiv 
\mathbf{m}_{A}\cdot \mathbf{\hat{\sigma}}_{A}$, $\hat{\sigma}_{B}\equiv 
\mathbf{m}_{B}\cdot \mathbf{\hat{\sigma}}_{B}$ yield possible outcomes $%
\sigma _{A}$, $\sigma _{B}=\pm 1$, for arbitrary axes $\mathbf{m}_{A}$, $%
\mathbf{m}_{B}$. Over an ensemble represented by the singlet state%
\begin{equation*}
\left\vert \Psi \right\rangle =\frac{1}{\surd 2}\left( \left\vert
+-\right\rangle -\left\vert -+\right\rangle \right)
\end{equation*}%
Bell's theorem \cite{Bell64} shows that to reproduce the quantum correlation%
\begin{equation*}
\left\langle \Psi \right\vert \hat{\sigma}_{A}\hat{\sigma}_{B}\left\vert
\Psi \right\rangle =-\mathbf{m}_{A}\cdot \mathbf{m}_{B}=-\cos \theta _{AB}
\end{equation*}%
between $A$ and $B$, any hidden-variables theory must take the nonlocal form%
\begin{equation}
\sigma _{A}=\sigma _{A}(\mathbf{m}_{A},\mathbf{m}_{B},\lambda
),\;\;\;\;\sigma _{B}=\sigma _{B}(\mathbf{m}_{A},\mathbf{m}_{B},\lambda )
\label{dethv2}
\end{equation}%
in order to obtain%
\begin{equation*}
\left\langle \sigma _{A}\sigma _{B}\right\rangle _{\mathrm{eq}}=\int
d\lambda \ \rho _{\mathrm{eq}}(\lambda )\sigma _{A}\sigma _{B}=-\cos \theta
_{AB}
\end{equation*}%
for some distribution $\rho _{\mathrm{eq}}(\lambda )$ of hidden variables.
The individual outcomes $\sigma _{A}$, $\sigma _{B}$ at $A$ and $B$ must
depend nonlocally on the distant settings.\footnote{%
More precisely, there must be a nonlocal dependence in at least one
direction.} And yet, the marginal statistics of outcomes at $A$ and $B$ do
not depend on the distant settings.

This `washing out' of nonlocality is a peculiarity of equilibrium.
Generically, for an arbitrary distribution $\rho (\lambda )\neq \rho _{%
\mathrm{eq}}(\lambda )$ -- but retaining the same nonlocal mappings (\ref%
{dethv2}) from individual conditions $\mathbf{m}_{A}$, $\mathbf{m}_{B}$, $%
\lambda $ to individual outcomes $\sigma _{A}$, $\sigma _{B}$ -- the
marginal statistics at one wing do depend instantaneously on the choice of
measurement axis at the distant wing. This is true not only in pilot-wave
theory \cite{PLA2}, but in any deterministic hidden-variables theory \cite%
{PLAc,Cracow}.

Thus, the locality property of quantum theory -- that entangled states
cannot be used for nonlocal signalling -- is a contingent feature of quantum
equilibrium, and is generically violated for nonequilibrium ensembles.

One might think that, in nonequilibrium, nonlocal signals would necessarily
create causal paradoxes. However, these can be evaded by modifying the
causal structure of spacetime. Specifically, one may assume that in
nonequilibrium there is a preferred foliation by spacelike hypersurfaces,
labelled by a time parameter $t$ that defines a fundamental causal sequence,
as in fact is the case in pilot-wave field theory (see section 2.4 below).
Nonlocal, nonequilibrium signals then define an absolute simultaneity.%
\footnote{%
On this view, `back-in-time' effects created by Lorentz boosts are
fictitious, because moving clocks are incorrectly synchronised if one
assumes isotropy of the speed of light in all frames \cite{AVth,AVbook,BH84}.%
}

In equilibrium, the underlying details of the hidden-variable dynamics are
washed out. The resulting `erasure of information' leads to the \textit{%
emergent} property of statistical locality, which is contingent on $\rho
(\lambda )=\rho _{\mathrm{eq}}(\lambda )$. An analogy may be drawn with the
state of classical thermodynamic heat death (or global thermal equilibrium),
in which the lack of differences of temperature makes it impossible to
convert heat into work \cite{AVth,PLA1,AV96}. Such a limitation is an
artifact of the state, not a fundamental law of physics.

The analogy with thermal equilibrium goes further and deeper. As noted in
the Introduction, thermal equilibrium has the remarkable effect of erasing
information about the past. For a gas in thermal equilibrium, the
macroscopic state contains no memory of how the state was formed. Further,
the equilibrium state may be described by a small number of parameters, and
its statistical properties are completely specified by the partition
function $Z$. These remarkable features of thermal equilibrium may be
summarised as `memory loss' and `information compression'. Similar features
emerge in quantum equilibrium.

In quantum theory, once a quantum state has been prepared, the predicted
probabilities carry no trace of \textit{how} the state was prepared. That
this is indeed remarkable has been emphasised by Peres \cite{Per}, who
adopts a postulate of `statistical determinism', according to which quantum
probabilities for a pure state do not depend on the details of the
preparation procedure.\footnote{%
This is a component of Peres' postulate A (ref. \cite{Per}, p. 30). As Peres
notes (p. 31), `\textit{all the past history of the selected quantum systems
becomes irrelevant}' (italics in original).} Peres also points out another
remarkable fact, that for mixed states a density operator $\hat{\rho}$ may
be prepared in an infinite number of macroscopically-distinct ways, and yet
no information distinguishing these preparations may be recovered from $\hat{%
\rho}$, which contains all the statistical information about the prepared
ensemble.\footnote{%
See ref. \cite{Per}, p. 75. This property is expressed in Peres' final
postulate K (p. 76).} The quantum probabilities for all possible quantum
measurements are described by the operator $\hat{\rho}$, and may be encoded
into a small number $K=N^{2}$ of `fiducial probabilities', where $N$ is the
(quantum) dimension of the system \cite{Hardy}.

Thus, there is a quantum `memory loss' with respect to state preparation,
analogous to the memory loss in equilibrium thermodynamics. Further, there
is a quantum `information compression', in which the statistics of outcomes
may be encoded into a small number $K$ of parameters and described by a
single mathematical object $\hat{\rho}$, analogous to the parameters $V$, $T$
(labelling thermodynamic states) and the partition function $Z$.

These features of quantum theory are -- from a hidden-variables perspective
-- contingencies of quantum equilibrium $\rho (\lambda )=\rho _{\mathrm{eq}%
}(\lambda )$. For nonequilibrium ensembles, the quantum state alone is not
sufficient to specify the probabilities for outcomes of quantum
measurements, as we saw in Sect. 2.1. In other words, $\hat{\rho}$ is
generally not a complete description, and additional details about the past
need not be erased. Different preparations of the same density operator $%
\hat{\rho}$ can be distinguished in nonequilibrium. As noted by Peres \cite%
{Per}, the ability to do so leads to nonlocal signalling. Measurements at
one wing of an entangled pure state may be regarded as preparing a mixed
state at the other distant wing, and as we have already noted, in
nonequilibrium the marginal statistics at the distant wing will generally
depend on what measurements were performed far away -- or equivalently, on
how the mixed state was prepared \cite{AVcont}. Thus, the reduced density
matrix at the distant wing is \textit{not} a complete description, and
information about the preparation is not lost.

The incompleteness of the reduced density matrix will be crucial in
understanding how Hawking information loss may be avoided, since a key
assumption of Hawking's argument is that the reduced density matrix in the
exterior region provides a complete description post-evaporation.

Note also that, in pilot-wave theory, the true decomposition of the density
operator cannot be determined in equilibrium, but the fundamental dynamics
does depend on it, and in nonequilibrium the true decomposition would be
apparent from the detailed behaviour of the trajectories.

It may be shown that the number $K$ of fiducial probabilities is larger than 
$N^{2}$ in nonequilibrium, an effect which may be traced to the extra
information about hidden variables that is revealed in nonequilibrium \cite%
{MVW}. Similar effects occur, of course, in thermal nonequilibrium, where
information about microscopic variables is no longer screened off, resulting
in a breakdown of what we have called memory loss and information
compression.

The fundamental message is that, in general, \textit{equilibrium erases
information}. In particular, quantum equilibrium erases information about
hidden variables (or de Broglie-Bohm trajectories), about nonlocal
interactions, and about how the system was prepared.

\subsection{The Role of Gravitation}

In non-gravitational physics, quantum equilibrium appears to be stable, in
the sense of being preserved in time under standard processes and
interactions. The Born rule continues to hold, for instance, in high-energy
collisions (as probed by scattering cross-sections). An example of a
hidden-variables theory with a stable quantum equilibrium state at high
energies is provided by the pilot-wave theory of fields in flat spacetime 
\cite{Bohm,AVth,Holl,BandH,AVbook,AV96,BH84,Kal85,BHK,HollPR,Kal94}. For a
scalar field $\phi $ (for example) one may write quantum field theory in the
functional Schr\"{o}dinger picture, in terms of a wave functional $\Psi
\lbrack \phi ,t]$, and one may assume that the velocity $\partial \phi
(x,t)/\partial t$ of the actual field configuration $\phi (x,t)$ is given by
the functional derivative $\delta S/\delta \phi (x)$ of the phase $S$ of $%
\Psi $ (or more generally, by the ratio of the quantum probability current $J
$ in field configuration space to the quantum probability density $|\Psi
|^{2}$). This is the natural generalisation of pilot-wave dynamics to
continuous degrees of freedom. The configuration $q(t)$ is now the field
configuration $\phi (x,t)$, and the wave function $\psi (q,t)$ is now the
wave functional $\Psi \lbrack \phi ,t]$. (The construction requires a
fundamental time parameter $t$, with respect to which nonlocal effects occur
instantaneously \cite{BH84,AVth}.) Standard quantum field theory, together
with ordinary local Lorentz symmetry, is recovered for ensembles of fields
in quantum equilibrium, that is, for fields distributed according to $P[\phi
,t]=|\Psi \lbrack \phi ,t]|^{2}$. As usual, this equilibrium distribution
need be given at some initial time only. Quantum equilibrium is therefore
stable with respect to high-energy collisions, at the energies probed so
far, in at least one hidden-variables theory.

Pilot-wave field theory may be readily extended to a curved spacetime that
is globally hyperbolic \cite{AVbook}. Let us sketch the construction. Any
globally hyperbolic spacetime may be foliated (in general nonuniquely) by
spacelike hypersurfaces $\Sigma $ labelled by a global time function $t$ 
\cite{HE}. The classical spacetime metric may be written as%
\begin{equation*}
d\tau ^{2}=\,^{(4)}g_{\mu \nu }dx^{\mu }dx^{\nu
}=N^{2}dt^{2}-g_{ij}dx^{i}dx^{j}
\end{equation*}%
We have set the shift vector $N^{i}=0$, so that lines $x^{i}=\mathrm{const}.$
are normal to $\Sigma $ (as may always be done as long as the lines $x^{i}=%
\mathrm{const}.$ do not run into singularities). The lapse function $%
N(x^{i},t)$ measures the proper time lapse normal to $\Sigma $ per unit of
coordinate time $t$. Now, restricting ourselves to a massless and
minimally-coupled scalar field $\phi $, the Lagrangian density $\mathcal{L}=%
\frac{1}{2}\sqrt{-\,^{(4)}g}\,^{(4)}g^{\mu \nu }\partial _{\mu }\phi
\partial _{\nu }\phi $ (with action $\int \int dtd^{3}x\ \mathcal{L}$) may
be written as%
\begin{equation*}
\mathcal{L}=\frac{1}{2}N\sqrt{g}\left( \frac{\dot{\phi}^{2}}{N^{2}}%
-g^{ij}\partial _{i}\phi \partial _{j}\phi \right)
\end{equation*}%
which implies a canonical momentum density $\pi =\partial \mathcal{L}%
/\partial \dot{\phi}=(\sqrt{g}/N)\dot{\phi}$ and a Hamiltonian%
\begin{equation*}
H=\int d^{3}x\;\frac{1}{2}N\sqrt{g}\left( \frac{\pi ^{2}}{g}+g^{ij}\partial
_{i}\phi \partial _{j}\phi \right)
\end{equation*}%
The wave functional $\Psi \lbrack \phi ,t]$ then satisfies the Schr\"{o}%
dinger equation (with units $\hbar =c=1$)%
\begin{equation}
i\frac{\partial \Psi }{\partial t}=\int d^{3}x\;\frac{1}{2}N\sqrt{g}\left( -%
\frac{1}{g}\frac{\delta ^{2}}{\delta \phi ^{2}}+g^{ij}\partial _{i}\phi
\partial _{j}\phi \right) \Psi  \label{Sch2}
\end{equation}%
which implies the continuity equation%
\begin{equation}
\frac{\partial \left\vert \Psi \right\vert ^{2}}{\partial t}+\int d^{3}x\;%
\frac{\delta }{\delta \phi }\left( \left\vert \Psi \right\vert ^{2}\frac{N}{%
\sqrt{g}}\frac{\delta S}{\delta \phi }\right) =0  \label{conta2}
\end{equation}%
and a de Broglie velocity field%
\begin{equation}
\frac{\partial \phi }{\partial t}=\frac{N}{\sqrt{g}}\frac{\delta S}{\delta
\phi }  \label{deB2}
\end{equation}%
(where $\Psi =\left\vert \Psi \right\vert e^{iS}$). According to (\ref{deB2}%
), the field velocity $\dot{\phi}$ at a point $x$ on the hypersurface $%
\Sigma $ depends instantaneously (with respect to $t$) on field values at
distant points $x%
{\acute{}}%
\neq x$ on $\Sigma $, if the wave functional is entangled with respect to
the fields at those points. To ensure physical consistency we assume, as in
the flat case, that the theory is constructed using a preferred foliation
associated with a specific lapse function $N(x^{i},t)$ (which then plays the
role of an additional physical field affecting the rate of macroscopic
clocks \cite{AVbook}). As before, (\ref{Sch2}) and (\ref{deB2}) determine
the evolution $\phi (x,t)$ of an individual field (given the initial
configuration $\phi (x,0)$ and wave functional $\Psi \lbrack \phi ,0]$). The
time evolution of an arbitrary ensemble distribution $P[\phi ,t]$ is then
given by%
\begin{equation}
\frac{\partial P}{\partial t}+\int d^{3}x\;\frac{\delta }{\delta \phi }%
\left( P\frac{N}{\sqrt{g}}\frac{\delta S}{\delta \phi }\right) =0
\label{contb2}
\end{equation}%
Comparing (\ref{conta2}) and (\ref{contb2}), it follows as usual that $%
P[\phi ,t]=\left\vert \Psi \lbrack \phi ,t]\right\vert ^{2}$ is an
equilibrium state. And in equilibrium, one may ignore the details of the
trajectories defined by (\ref{deB2}), and consider only the probabilities
obtained from the (modulus-squared of the) wave functional governed by (\ref%
{Sch2}), which will agree with ordinary quantum field theory on curved
spacetime. Thus, again, even in the presence of gravitation, the quantum
equilibrium state is stable in at least one hidden-variables theory --
provided the spacetime is globally hyperbolic.\footnote{%
We are of course glossing over the question of the rigorous definition of
the functional Schr\"{o}dinger equation, which is used extensively in (for
example) cosmological inflationary models. We are implicitly assuming some
sort of regularisation, such as an analytical continuation of the number of
space dimensions away from 3 \cite{Guv89}.}

In contrast, for a non-globally-hyperbolic spacetime -- such as that
generated by the formation and (complete) evaporation of a black hole --
even standard quantum field theory has been developed to only a very limited
degree. And a pilot-wave analogue has not been developed at all.

According to Hawking's argument, an initial pure state $|\Psi \rangle $ may
be defined on some (global) initial spacelike hypersurface $\Sigma _{1}$,
before the hole forms (treating spacetime as a classical background). Once
the horizon has formed, and evaporation begins, the Hilbert space may be
written as a product $\mathcal{H}_{\mathrm{int}}\otimes \mathcal{H}_{\mathrm{%
ext}}$ over the degrees of freedom interior and exterior to the hole. On a
hypersurface $\Sigma _{2}$ that crosses the horizon, the quantum state of
the system is still the pure state $|\Psi \rangle $ (in the Heisenberg
picture). However, the quantum state in the exterior region is mixed, and is
represented by the reduced density operator $\hat{\rho}_{\mathrm{ext}}=Tr_{%
\mathrm{int}}(|\Psi \rangle \langle \Psi |)$, obtained by tracing over the
interior degrees of freedom. After the black hole has evaporated, the mixed
state $\hat{\rho}_{\mathrm{ext}}$ is the state of the whole system, defined
on a final hypersurface $\Sigma _{3}$.

To construct quantum field theory on such a spacetime is not
straightforward. The standard quantisation procedure uses canonical
commutation relations on a Cauchy surface, so that the wave equation has a
well-posed initial value formulation \cite{Wald94}; it relies on quantising
a well-posed Hamiltonian dynamics for classical fields, and is therefore
applicable only to globally-hyperbolic spacetimes.\footnote{%
The derivation of Hawking radiance is based on quantum field theory on a
background globally-hyperbolic spacetime. Once the radiation rate has been
derived, it is assumed that the mass of the hole steadily decreases to
compensate, and assuming complete evaporation results in a
non-globally-hyperbolic spacetime. The pure-to-mixed transition is
intimately related to the failure of global hyperbolicity \cite{Wald94}.} An
algebraic approach to quantum field theory on non-globally-hyperbolic
spacetimes has been developed by Yurtsever \cite{Yurt94} and applied to
simple, flat (two-dimensional) examples. In this construction, it is
insufficient to specify the (algebraically-defined) quantum state on an
initial spacelike hypersurface; the state must be specified on the entire
spacetime, with boundary conditions at naked singularities (if any).

It remains to be seen if Yurtsever's approach can be extended to a
hidden-variables theory. Extant deterministic theories require a preferred
hypersurface along which nonlocality acts, as in equation (\ref{deB2}). Even
in flat spacetime, attempts to write down a fundamentally Lorentz-invariant
theory of (hidden) particle trajectories run into problems associated with
nonlocality, so that the quantum equilibrium distribution must be defined on
a preferred hypersurface \cite{Har92,BG,Detc,Myr}.\footnote{%
See, however, ref. \cite{DH} for an attempt to circumvent this.} In the
absence of any Cauchy hypersurface whatever, there might be a fundamental
difficulty in defining a quantum equilibrium state for a nonlocal
hidden-variables theory. In particular, the de Broglie-Bohm construction
depends on the existence of a local quantum probability current in
configuration space, and it is not clear that such a current will exist;
though this remains to be studied.\footnote{%
Note that the de Broglie-Bohm version of canonical quantum gravity (which
has been applied to Hawking evaporation \cite{Aca}) is not relevant here,
because the canonical formalism -- in which general relativity is written as
the theory of a 3-geometry evolving in time (with respect to an arbitrary
foliation) -- has physical significance only if spacetime is globally
hyperbolic.}

In the usual pilot-wave dynamics of a single system, there is a
deterministically evolving, well-defined wave function (or functional) at
all times, generating a velocity field in configuration space. Mixed states
may, as we have mentioned, be interpreted in terms of ordinary statistical
mixtures of such (physically real) wave functions. A hypothetical transition
from a pure to a mixed state, however, would in de Broglie-Bohm theory
amount to a denial of the deterministic evolution of the pilot wave.
Possibly, one could extend ordinary pilot-wave theory by introducing a
stochastic element in the evolution of the wave function, to account for
pure-to-mixed transitions; or otherwise define a local current by a more
general prescription that is not tied to the wave function. If this could be
done, the resulting theory might exhibit an equilibrium distribution that
agrees with quantum probabilities throughout the transition.\footnote{%
Maroney \cite{Mar} considers a de Broglie-Bohm-type theory in which the
density operator plays the role of a physical guiding field for individual
systems.} Otherwise, if one begins with a pure quantum state $\Psi $ and a
quantum equilibrium distribution $P=|\Psi |^{2}$ at some initial time, then
in the absence of a detailed theory there is no guarantee that upon
transition to a mixed state $\hat{\rho}$ the fields will end with the
quantum distribution $\langle \phi |\hat{\rho}|\phi \rangle $.

At present, then, there is no known example of a hidden-variables theory
exhibiting a stable state of quantum equilibrium on a
non-globally-hyperbolic spacetime or in the presence of a pure-to-mixed
transition. And even if it proves possible to construct such a theory, one
might still consider the possibility that in such situations -- and in
particular in the spacetime generated by the formation and evaporation of a
black hole -- the quantum equilibrium state becomes unstable.

From a hidden-variables perspective, it is natural to ask if there exist
some physical processes in which quantum nonequilibrium can be generated
from an initial equilibrium state. Possibly, the quantum equilibrium state
becomes unstable at very high (hitherto-unprobed) energies or in strong
gravitational fields, where quantum theory might break down. However, rather
than simply postulating arbitrary new effects, one would like theoretical
reasons for why nonequilibrium might be generated in specific conditions.

Gravitation seems the natural arena in which to search for such effects,
since it is not really known how gravitation fits in with quantum theory.
Further, Hawking's argument for information loss already suggests that
gravitation may force a significant departure from the usual quantum
formalism.

We shall now argue that hidden-variable degrees of freedom can be `unfrozen'
by gravitational effects. In particular, we suggest that black holes can
throw systems out of quantum equilibrium, if those systems are entangled
with other systems that have fallen behind the event horizon of the black
hole. Hidden variables provide an additional entropy reservoir that is
usually filled (in equilibrium with $S_{\mathrm{hv}}=0$), and remains filled
in all known processes. And because details of the hidden variables are
erased in equilibrium, they may be ignored in ordinary (quantum) physics.
However, if information is allowed to flow into and out of this reservoir of
extra degrees of freedom (so that $S_{\mathrm{hv}}$ changes), then new
phenomena can arise that lie outside the scope of quantum physics.

\section{A Thought Experiment with Black Holes and Entangled States}

Consider the following thought experiment. Outside the event horizon of a
black hole, a source emits EPR-pairs -- specifically, pairs of entangled
two-state systems (labelled $A$ and $B$) prepared in the (pure) singlet state%
\begin{equation*}
\left\vert \Psi \right\rangle =\frac{1}{\sqrt{2}}\left( \left\vert
+-\right\rangle -\left\vert -+\right\rangle \right)
\end{equation*}%
The states $\left\vert +\right\rangle $, $\left\vert -\right\rangle $ could
for example be orthogonal states of spin, polarisation, momentum or energy.
The ensemble of EPR-pairs is represented by a density operator $\hat{\rho}^{%
\mathrm{in}}=\left\vert \Psi \right\rangle \langle \Psi |$, and the von
Neumann entropy is zero,%
\begin{equation*}
S_{\mathrm{vonN}}^{\mathrm{in}}=-Tr(\hat{\rho}^{\mathrm{in}}\ln \hat{\rho}^{%
\mathrm{in}})=0
\end{equation*}

Now, if one half of each pair falls behind the event horizon, then the
reduced density operator for the ensemble of systems in the exterior region
will be%
\begin{equation}
\hat{\rho}_{\mathrm{ext}}=\frac{1}{2}\left\vert +\right\rangle \langle +|+%
\frac{1}{2}\left\vert -\right\rangle \langle -|  \label{Mixed}
\end{equation}%
At this stage, the total state is still pure, and the mixed state in the
exterior region is merely the result of tracing over degrees of freedom that
have fallen behind the horizon. However, if we now wait for the black hole
to evaporate completely, the interior region disappears and the remaining
two-state systems are left in a strictly mixed state%
\begin{equation}
\hat{\rho}^{\mathrm{out}}=\hat{\rho}_{\mathrm{ext}}  \label{rhoout}
\end{equation}%
Of course, these systems will now be accompanied by a bath of thermal
radiation. But if we assume that the systems in the exterior region were
carried far away from the hole, and (or) appropriately isolated, they will
not have interacted with the emerging Hawking radiation: thus, they will not
be entangled with any of the emitted particles, and their quantum state will
indeed be just (\ref{rhoout}), with a von Neumann entropy%
\begin{equation*}
S_{\mathrm{vonN}}^{\mathrm{out}}=-Tr(\hat{\rho}^{\mathrm{out}}\ln \hat{\rho}%
^{\mathrm{out}})=\ln 2
\end{equation*}%
The increase in entropy by $\ln 2$ quantifies the `information loss'
associated with the transition from a pure to a mixed quantum state.

We now make the following \textit{hypothesis}: that during the pure-to-mixed
transition, the ensemble of systems outside the horizon evolves away from
quantum equilibrium, so that the total entropy $S_{\mathrm{hv}}+S_{\mathrm{%
vonN}}$ is conserved.

According to this hypothesis, the ensemble of systems post-evaporation will
be equivalent to an ensemble that was subjected to a preparation procedure
represented by (\ref{Mixed}), but whose distribution of hidden variables
differs from quantum equilibrium, $\rho ^{\mathrm{out}}(\lambda )\neq \rho _{%
\mathrm{eq}}(\lambda )$, where $\rho ^{\mathrm{out}}(\lambda )$ is such that
the change in hidden-variable entropy balances the change in von Neumann
entropy:%
\begin{equation}
S_{\mathrm{hv}}^{\mathrm{out}}+S_{\mathrm{vonN}}^{\mathrm{out}}=S_{\mathrm{hv%
}}^{\mathrm{in}}+S_{\mathrm{vonN}}^{\mathrm{in}}  \label{ConsS}
\end{equation}%
If the initial quantum state is pure, and if the initial ensemble is in
quantum equilibrium, then $S_{\mathrm{vonN}}^{\mathrm{in}}=S_{\mathrm{hv}}^{%
\mathrm{in}}=0$ and 
\begin{equation}
S_{\mathrm{hv}}^{\mathrm{out}}=-\ln 2  \label{log2}
\end{equation}%
This provides a quantitative prediction for the amount of nonequilibrium
generated.

Note that in (\ref{ConsS}) we are comparing two very different kinds of
entropy, $S_{\mathrm{hv}}$ and $S_{\mathrm{vonN}}$. These may in fact be
related even in non-gravitational processes, as we shall discuss in section
5.

Further, the assumed conservation rule (\ref{ConsS}) is merely a simple
hypothesis. If a process of this type really does occur, one hopes that (\ref%
{ConsS}) will be correct at least as an order-of-magnitude estimate. In any
case, as we shall see in section 6, the estimated change (\ref{log2}) might
be susceptible to experimental test.

A detailed theory of the proposed process will not be attempted here.
However, intuitively we envisage the mechanism to be along these lines. The
entangled systems interior and exterior to the horizon are nonlocally
connected at the hidden-variable level (presumably along some unknown
spacelike hypersurface, at least prior to complete evaporation). As the
interior system approaches the singularity, something happens to it that we
do not currently understand, and this is communicated nonlocally to the
exterior system, causing the latter (over an ensemble) to fall out of
quantum equilibrium. It might be the case, for example, that in a black hole
the internal degrees of freedom close to the singularity are far from
quantum equilibrium, and that when the infalling system interacts with these
it communicates the nonequilibrium to the exterior system.

This last possibility may be illustrated with a simple model based on de
Broglie-Bohm theory. Consider two nonrelativistic particles moving in one
spatial dimension, with positions $x_{A}$ and $x_{B}$. Let their initial
wave function $\psi _{0}(x_{A},x_{B})$ be entangled, and assume that (over
an ensemble) the particles are in quantum equilibrium, with a joint
distribution $\left\vert \psi _{0}(x_{A},x_{B})\right\vert ^{2}$ of
positions. Now let particle $A$ interact locally with a third particle, with
position $y$ and initial wave function $\phi _{0}(y)$, via the Hamiltonian $%
\hat{H}=a\hat{y}\hat{p}_{A}$ (where $a$ is a coupling constant and $\hat{p}%
_{A}$ is conjugate to $\hat{x}_{A}$). It is straightforward to show that if $%
y$ has a nonequilibrium distribution, then provided $\psi _{0}(x_{A},x_{B})$
is entangled the local interaction between $y$ and $x_{A}$ drives the
marginal distribution of $x_{B}$ away from equilibrium. For let the coupling 
$a$ be so large that the Hamiltonians of the particles themselves may be
neglected. Then the wave function $\Psi (x_{A},x_{B},y,t)$ obeys the Schr%
\"{o}dinger equation%
\begin{equation}
i\frac{\partial \Psi }{\partial t}=-iay\frac{\partial \Psi }{\partial x_{A}}
\label{SchAB}
\end{equation}%
and the initial quantum state $\Psi _{0}(x_{A},x_{B},y)=\psi
_{0}(x_{A},x_{B})\phi _{0}(y)$ evolves into%
\begin{equation*}
\Psi (x_{A},x_{B},y,t)=\psi _{0}(x_{A}-ayt,x_{B})\phi _{0}(y)
\end{equation*}%
From (\ref{SchAB}) one obtains the continuity equation%
\begin{equation*}
\frac{\partial \left\vert \Psi \right\vert ^{2}}{\partial t}+ay\frac{%
\partial \left\vert \Psi \right\vert ^{2}}{\partial x_{A}}=0
\end{equation*}%
which implies that the (hidden-variable) positions have velocities $\dot{x}%
_{A}=ay$, $\dot{x}_{B}=0$, $\dot{y}=0$ and trajectories%
\begin{equation*}
x_{A}(t)=x_{A}(0)+ay(0)t,\;\;\;x_{B}(t)=x_{B}(0),\;\;\;y(t)=y(0)
\end{equation*}%
An arbitrary distribution $P(x_{A},x_{B},y,t)$ then evolves according to the
same continuity equation%
\begin{equation*}
\frac{\partial P}{\partial t}+ay\frac{\partial P}{\partial x_{A}}=0
\end{equation*}%
and an initial distribution $P_{0}(x_{A},x_{B},y)=\left\vert \psi
_{0}(x_{A},x_{B})\right\vert ^{2}\pi _{0}(y)$ with $\pi _{0}(y)\neq
\left\vert \phi _{0}(y)\right\vert ^{2}$ (that is, with $x_{A},x_{B}$ in
equilibrium and $y$ in nonequilibrium) evolves into%
\begin{equation*}
P(x_{A},x_{B},y,t)=\left\vert \psi _{0}(x_{A}-ayt,x_{B})\right\vert ^{2}\pi
_{0}(y)
\end{equation*}%
This is to be compared with the equilibrium result%
\begin{equation*}
P_{\mathrm{eq}}(x_{A},x_{B},y,t)=\left\vert \psi
_{0}(x_{A}-ayt,x_{B})\right\vert ^{2}\left\vert \phi _{0}(y)\right\vert ^{2}
\end{equation*}%
In particular, for $\pi _{0}(y)\neq \left\vert \phi _{0}(y)\right\vert ^{2}$
the marginal distribution $\rho (x_{B},t)$ at $B$%
\begin{equation}
\rho (x_{B},t)=\int dx_{A}\left( \int dy\ \left\vert \psi
_{0}(x_{A}-ayt,x_{B})\right\vert ^{2}\pi _{0}(y)\right)  \label{rhoB}
\end{equation}%
will generally differ from the equilibrium marginal%
\begin{equation}
\rho _{\mathrm{eq}}(x_{B},t)=\int dx_{A}\left( \int dy\ \left\vert \psi
_{0}(x_{A}-ayt,x_{B})\right\vert ^{2}\left\vert \phi _{0}(y)\right\vert
^{2}\right)  \label{rhoeqB}
\end{equation}%
(Note, however, that for a product state $\psi _{0}(x_{A},x_{B})=\alpha
(x_{A})\beta (x_{B})$ both (\ref{rhoB}) and (\ref{rhoeqB}) reduce to $%
\left\vert \beta (x_{B})\right\vert ^{2}$ (since $\int dx_{A}\ \left\vert
\alpha (x_{A}-ayt)\right\vert ^{2}=1$), and the system at $B$ remains in
equilibrium.) Thus, an entangled state of ordinary quantum systems $A$ and $%
B $ can provide a channel for the nonlocal propagation of nonequilibrium
from $A$ to $B$, if we allow a nonequilibrium system to interact locally
with $A$.

In practice, each infalling system is likely to undergo different
interactions with the interior, and the ensemble distribution for the
exterior systems will be obtained by averaging over such interactions. It
might then be thought that even if there were nonequilibrum degrees of
freedom behind the horizon, such averaging would result in an equilibrium
distribution for the exterior ensemble. However, a `thermodynamic'
constraint of the form (\ref{ConsS}) (if such exists) would prevent this.

Because we do not have a detailed theory, we do not know along which
spacelike hypersurface the nonlocality acts. Thus, we cannot say how long an
experimenter outside the horizon would have to wait in order to see the
exterior systems fall out of quantum equilibrium. However, it seems unlikely
that the process should occur only in the final stages of evaporation. For
consider a black hole that had been formed entirely from the collapse of one
half of an EPR-ensemble. In the late stages of evaporation, the other half
of the EPR-ensemble (assumed to be far away from the hole and shielded from
the outgoing radiation) will be essentially mixed, because a Planck-sized
remnant cannot store the quantum information required to maintain overall
purity (without having an unreasonably large number of internal states). If
information loss is to be avoided, the nonequilibrium transition must take
place earlier, even when the hole is macroscopic.

We shall therefore assume that the process takes place whenever one half of
an EPR-pair falls behind the event horizon, even for a macroscopic black
hole. And, we shall assume that the process occurs in a `reasonably
accessible time', with respect to a distant experimenter.

Note that (\ref{log2}) entails a large deviation from quantum probabilities.
For example, let $\left\vert \pm \right\rangle $ be momentum states $%
\left\vert \pm \mathbf{p}\right\rangle $ of a free (nonrelativistic)
particle of energy $E=p^{2}/2m$ in some normalisation volume $V$. In
pilot-wave theory, the quantum state (\ref{Mixed}) consists of a statistical
mixture of objectively-existing wave functions, which we take to be (for
example)%
\begin{equation*}
\psi _{+}(\mathbf{x},t)=\frac{1}{\sqrt{V}}e^{i\mathbf{p}\cdot \mathbf{x}%
}e^{-iEt},\;\;\;\;\psi _{-}(\mathbf{x},t)=\frac{1}{\sqrt{V}}e^{-i\mathbf{p}%
\cdot \mathbf{x}}e^{-iEt}
\end{equation*}%
As discussed in section 2.1, for each pure subensemble -- with wave function 
$\psi _{\pm }$ and distribution $\rho _{\pm }$ -- we may define a
hidden-variable entropy%
\begin{equation*}
S_{\mathrm{hv}}^{\pm }=-\int d^{3}\mathbf{x}\;\rho _{\pm }\ln (\rho _{\pm
}/\left\vert \psi _{\pm }\right\vert ^{2})
\end{equation*}%
and for the whole ensemble the (mean) hidden-variable entropy is $S_{\mathrm{%
hv}}=\frac{1}{2}S_{\mathrm{hv}}^{+}+\frac{1}{2}S_{\mathrm{hv}}^{-}$ (where
in quantum equilibrium $S_{\mathrm{hv}}=0$, while out of equilibrium $S_{%
\mathrm{hv}}<0$). Taking for simplicity $\rho _{+}=\rho _{-}=\rho $, (\ref%
{log2}) becomes%
\begin{equation}
\int d^{3}\mathbf{x}\;\rho \ln (\rho V)=\ln 2  \label{int=ln2}
\end{equation}%
A simple distribution $\rho $ satisfying this equality is obtained by
dividing the volume $V$ into two halves, and setting $\rho =0$ in one half
and $\rho =2/V$ in the other. There are, of course, an infinite number of
nonequilibrium distributions satisfying (\ref{int=ln2}). Possibly, a
detailed theory with a specific mechanism for generating nonequilibrium
would yield a definite prediction for the final-state distribution $\rho $;
this remains to be seen. In any case, it is clear that the required degree
of nonequilibrium is rather large.

According to our argument, then, throwing one half of an EPR-pair into a
black hole will drive the hidden variables for the remaining half away from
quantum equilibrium (over an ensemble). Subsequent experiments with these
systems will then yield non-standard (non-quantum) probabilities.

This process may be viewed in terms of a nonlocal leakage of information
from behind the horizon. For example, let the initial state consist of a
pair of photons whose polarisations are entangled. Allowing one photon to
fall behind the horizon, let the other now be directed towards a polariser,
followed by a second polariser at angle $\Theta $ with respect to the first.
According to quantum theory, any photons passing through the first polariser
are prepared in a quantum state of definite (linear) polarisation, and for
these photons the probability of transmission through the second polariser
will be%
\begin{equation}
p_{\mathrm{eq}}^{+}(\Theta )=\cos ^{2}\Theta  \label{cos2}
\end{equation}%
Whatever the state of photons entering the first polariser, the fraction
transmitted through the second must vary as $\cos ^{2}\Theta $ as the angle $%
\Theta $ is varied. But we saw in section 2.2 that, in any deterministic
hidden-variables theory, a nonequilibrium distribution $\rho (\lambda )\neq
\rho _{\mathrm{eq}}(\lambda )$ will generically imply a non-sinusoidal
transmission probability (\ref{NQp}) or%
\begin{equation}
p^{+}(\Theta )\neq \cos ^{2}\Theta  \label{notcos2}
\end{equation}%
(where here the values $\sigma =\pm 1$ of the observable $\hat{\sigma}=%
\mathbf{m}\cdot \mathbf{\hat{\sigma}}$ correspond respectively to
polarisation parallel or perpendicular to an axis in 3-space with angle $%
\Theta =\theta /2$). In effect, the non-sinusoidal distribution $p^{\pm
}(\Theta )$ would contain information from behind the horizon. Heuristically
speaking, the usual one-way membrane is `pierced' by the entangled state,
which provides a channel along which information may leak out, in the form
of non-standard polarisation probabilities.\footnote{%
We are assuming, of course, that nonequilibrium is produced in the exterior
photon before it enters the first polariser (while it is still entangled
with the photon behind the horizon). Note also that there is no particular
reason why passage through the first polariser should cause the photon to
relax back to equilibrium (though this might happen in some hidden-variables
theories). For example, in de Broglie-Bohm theory, the division of a
nonequilibrium particle ensemble by a Stern-Gerlach spin measurement
generally results in nonequilibrium within each of the final (separated)
wave packets.}

The above discussion for the singlet state is easily generalised. Any pure
state for the pair of two-state systems may be written by Schmidt
decomposition in the form%
\begin{equation}
\left\vert \Psi \right\rangle =\sqrt{p_{1}}\left\vert A+,B+\right\rangle +%
\sqrt{p_{2}}\left\vert A-,B-\right\rangle  \label{Psi}
\end{equation}%
where $p_{1}$, $p_{2}$ are real and non-negative with $p_{1}+p_{2}=1$, and $%
\left\vert A\pm \right\rangle $ and $\left\vert B\pm \right\rangle $ are
orthonormal bases for systems $A$ and $B$ respectively. The reduced density
operator (for either system, suppressing the labels $A$, $B$ in the kets)
takes the form%
\begin{equation*}
\hat{\rho}_{A,B}=p_{1}\left\vert +\right\rangle \langle +|+p_{2}\left\vert
-\right\rangle \langle -|
\end{equation*}%
with a von Neumann entropy%
\begin{equation}
S_{\mathrm{vonN}}^{A,B}=-p_{1}\ln p_{1}-p_{2}\ln p_{2}  \label{vonNAB}
\end{equation}%
If one half of each pair falls behind the horizon, the conservation
hypothesis (\ref{ConsS}) implies that the external ensemble will acquire a
(nonequilibrium) hidden-variable entropy given by%
\begin{equation}
S_{\mathrm{hv}}^{\mathrm{out}}=p_{1}\ln p_{1}+p_{2}\ln p_{2}
\end{equation}%
Note that for any pure state (\ref{Psi}), the amount of nonequilibrium
generated -- that is, the value of $\left\vert S_{\mathrm{hv}}^{\mathrm{out}%
}\right\vert $ -- does not depend on which half ($A$ or $B$) of the ensemble
falls behind the horizon. This is because the von Neumann entropy (\ref%
{vonNAB}) for the reduced state is the same at each wing.

The entanglement entropy (\ref{vonNAB}) is a good measure of entanglement
for the state (\ref{Psi}). It is equal to $\ln 2$ times the number of
maximally-entangled states -- or Bell states -- into which $\left\vert \Psi
\right\rangle $ may be interconverted (by local operations at each wing and
classical communication alone) \cite{NC}. The amount of nonequilibrium
generated is therefore in direct proportion to the degree of entanglement of
the initial state, $\left\vert S_{\mathrm{hv}}^{\mathrm{out}}\right\vert =S_{%
\mathrm{vonN}}^{A,B}$. The nonequilibrium generated may also be related to
the maximal violation of Bell's inequality obtainable from quantum
measurements at each wing of the state (\ref{Psi}). In appropriate units,
the eigenvalues of the states $\left\vert A\pm \right\rangle $, $\left\vert
B\pm \right\rangle $ take values $a$, $b=\pm 1$, and in any \textit{local}
hidden-variables theory the expectation value $E(a,b)$ of the product $ab$
satisfies a Bell inequality (in the form due to Clauser \textit{et al}. \cite%
{CHSH69})%
\begin{equation}
\left\vert E(a,b)+E(a%
{\acute{}}%
,b)+E(a,b%
{\acute{}}%
)-E(a%
{\acute{}}%
,b%
{\acute{}}%
)\right\vert \leq 2  \label{CHSH}
\end{equation}%
where the primes denote a different orthonormal basis (corresponding to a
different measurement setting, for example measurement of polarisation along
a different axis). In quantum theory, for a state (\ref{Psi}) the left hand
side of (\ref{CHSH}) attains a maximum value \cite{GPR}%
\begin{equation}
f_{\max }=2\sqrt{1+4p_{1}p_{2}}  \label{Max}
\end{equation}%
Both (\ref{vonNAB}) and (\ref{Max}) are minimised by $p_{1}=0$ (or $p_{2}=0$%
) and maximised by $p_{1}=1/2$; further, both are monotonically increasing
functions of $p_{1}$ (or $p_{2}$) on the interval $(0,1/2)$. Thus, the
amount of nonequilibrium generated $\left\vert S_{\mathrm{hv}}^{\mathrm{out}%
}\right\vert =S_{\mathrm{vonN}}^{A,B}$ grows monotonically with the maximal
violation of Bell's inequality $f_{\max }$. This accords with our intuitive
picture of the exterior system being thrown out of equilibrium by nonlocal
effects from behind the horizon.

\section{Nonequilibrium Hawking Radiation}

As mentioned in the Introduction, outgoing modes for Hawking radiation are
entangled with ingoing modes. Mode-by-mode, then, the situation is
conceptually similar to the thought experiment just described, in which one
half of an EPR-pair falls behind the horizon. One may then similarly posit
an evolution away from quantum equilibrium for the outgoing radiation,
resulting in emitted particles whose statistical behaviour deviates from the
quantum formalism.

To illustrate the entanglement involved, for simplicity we shall first
consider the (rather artificial) case of the `eternal' black hole restricted
to two dimensions \cite{BD}. The standard Schwarzchild line element%
\begin{equation*}
d\tau ^{2}=\left( 1-\frac{2M}{r}\right) dt^{2}-\left( 1-\frac{2M}{r}\right)
^{-1}dr^{2}
\end{equation*}%
may be rewritten in terms of Kruskal null coordinates $\bar{u}$, $\bar{v}$ as%
\begin{equation*}
d\tau ^{2}=\frac{2M}{r}e^{-r/2M}d\bar{u}d\bar{v}
\end{equation*}%
Extending $\bar{u}$, $\bar{v}$ over the full range $(-\infty ,+\infty )$,
one obtains the maximally-extended Kruskal manifold. This contains two
asymptotically-flat regions I and II that are causally disconnected.

Quantising a massless scalar field $\phi $ on this spacetime, the wave
equation has natural basis modes $\propto e^{-i\omega \bar{u}}$, $%
e^{-i\omega \bar{v}}$ which are regular on the whole manifold. These have an
associated vacuum state $\left\vert 0\right\rangle _{\mathrm{K}}$ (the
Hartle-Hawking vacuum), where the modes have positive frequency with respect
to the time coordinate $\bar{t}=\frac{1}{2}(\bar{u}+\bar{v})$. By relating
these modes to modes defined in regions I and II only, the state $\left\vert
0\right\rangle _{\mathrm{K}}$ may be written in terms of particle states $%
\left\vert n_{k}\right\rangle _{\mathrm{I}}$ and $\left\vert
n_{k}\right\rangle _{\mathrm{II}}$ for regions I and II respectively:%
\begin{equation}
\left\vert 0\right\rangle _{\mathrm{K}}=\dprod\limits_{k}\frac{1}{\cosh
\alpha _{\omega }}\left( \sum_{n_{k}=0}^{\infty }e^{-4\pi Mn_{k}\omega
}\left\vert n_{k}\right\rangle _{\mathrm{I}}\otimes \left\vert
n_{k}\right\rangle _{\mathrm{II}}\right)  \label{0K}
\end{equation}%
where $\omega =\left\vert k\right\vert $ and $\tanh \alpha _{\omega }\equiv
e^{-4\pi M\omega }$.

While there is no entanglement between different modes in (\ref{0K}), each
mode is entangled between the causally-disconnected regions I and II. For
experiments performed in region I, all quantum probabilities are given by
the reduced density operator%
\begin{equation}
\hat{\rho}_{\mathrm{I}}=\dprod\limits_{k}\hat{\rho}_{\mathrm{I}}^{k}
\label{rhoI}
\end{equation}%
where for each mode%
\begin{equation}
\hat{\rho}_{\mathrm{I}}^{k}=\sum_{n_{k}=0}^{\infty }\frac{e^{-n_{k}\omega /T}%
}{Z_{k}}\left\vert n_{k}\right\rangle \left\langle n_{k}\right\vert
\label{rhoIk}
\end{equation}%
with $T=1/8\pi M$ and%
\begin{equation*}
Z_{k}=\sum_{n_{k}=0}^{\infty }e^{-n_{k}\omega /T}=(1-e^{-\omega /T})^{-1}
\end{equation*}

The state (\ref{rhoI}) corresponds to a thermal mixture at temperature $%
T=1/8\pi M$. The Hartle-Hawking vacuum $\left\vert 0\right\rangle _{\mathrm{K%
}}$ represents a black hole in thermal equilibrium with a bath of blackbody
radiation \cite{HH}. A more realistic choice of vacuum on the extended
Kruskal manifold, due to Unruh \cite{U76}, yields a purely outgoing thermal
flux.

The von Neumann entropy associated with $\hat{\rho}_{\mathrm{I}}$ is a sum%
\begin{equation}
S_{\mathrm{vonN}}^{\mathrm{I}}=\sum_{k}S_{\mathrm{vonN}}^{\mathrm{I,\ }k}
\label{vonNI}
\end{equation}%
of entropies of the individual modes%
\begin{equation*}
S_{\mathrm{vonN}}^{\mathrm{I,\ }k}=-Tr(\hat{\rho}_{\mathrm{I}}^{k}\ln \hat{%
\rho}_{\mathrm{I}}^{k})=-\sum_{n_{k}=0}^{\infty }p(n_{k})\ln p(n_{k})
\end{equation*}%
where $p(n_{k})=e^{-n_{k}\omega /T}/Z_{k}$. We have%
\begin{equation}
S_{\mathrm{vonN}}^{\mathrm{I,\ }k}=\frac{\omega }{T}(e^{\omega
/T}-1)^{-1}-\ln (1-e^{-\omega /T})  \label{vonNIk}
\end{equation}%
(The total entanglement entropy (\ref{vonNI}) diverges in the continuum
limit, and is rendered finite by an appropriate high-frequency cutoff.)

For the case of actual gravitational collapse, one begins with the Minkowski
vacuum $\left\vert 0\right\rangle _{\mathrm{M}}^{\mathrm{in}}$ in the
distant past (assuming the infalling matter to be arbitrarily tenuous at
arbitrarily early times). In the distant future, after the horizon has
formed, one may define an asymptotic Minkowski vacuum $\left\vert
0\right\rangle _{\mathrm{M}}^{\mathrm{out}}$. One finds that $\left\vert
0\right\rangle _{\mathrm{M}}^{\mathrm{in}}$ is a thermal mixture of particle
states defined with respect to $\left\vert 0\right\rangle _{\mathrm{M}}^{%
\mathrm{out}}$ (in the exterior region, outside the horizon). In other
words, the in-state is a thermal state with respect to the asymptotic
outgoing modes. The ingoing and outgoing field modes (representing pair
creation near the horizon) are entangled, and the reduced density operator
for the asymptotically Minkowski region is obtained by tracing over the
ingoing modes, resulting in a thermal mixture of outgoing radiation \cite{BD}%
.

Each mode $k$ (in a two-dimensional model) of the outgoing Hawking radiation
is represented by a thermal mixture $\hat{\rho}^{\mathrm{out,\ }k}$ of the
form (\ref{rhoIk}), with a von Neumann entropy $S_{\mathrm{vonN}}^{\mathrm{%
out,\ }k}$ given by (\ref{vonNIk}). Post-evaporation, the pure in-state
becomes a mixture%
\begin{equation}
\left\vert 0\right\rangle _{\mathrm{M}}^{\mathrm{in}}\rightarrow \hat{\rho}^{%
\mathrm{out}}=\tprod\limits_{k}\hat{\rho}^{\mathrm{out,\ }k}  \label{ptom}
\end{equation}

As in the thought experiment of section 4, we propose that during the
pure-to-mixed transition (\ref{ptom}), the emitted particles evolve away
from quantum equilibrium, by an amount such that the total entropy $S_{%
\mathrm{hv}}+S_{\mathrm{vonN}}$ is conserved.

An appropriate hidden-variable entropy $S_{\mathrm{hv}}$ may be defined in
pilot-wave field theory. Assuming the state $\hat{\rho}^{\mathrm{out}}$ to
be a thermal mixture of wave functionals%
\begin{equation*}
\Psi _{n_{k}n_{k%
{\acute{}}%
}....}[\phi ]=\langle \phi |n_{k}n_{k%
{\acute{}}%
}....\rangle
\end{equation*}%
for Fock states $|n_{k}n_{k%
{\acute{}}%
}....\rangle $, each pure subensemble with wave functional $\Psi _{n_{k}n_{k%
{\acute{}}%
}....}[\phi ]$ has a field distribution $P_{n_{k}n_{k%
{\acute{}}%
}....}^{\mathrm{out}}[\phi ]$ and a hidden-variable entropy%
\begin{equation*}
S_{\mathrm{hv}}^{\mathrm{out,\ }n_{k}n_{k%
{\acute{}}%
}....}=-\int D\phi \;P_{n_{k}n_{k%
{\acute{}}%
}....}^{\mathrm{out}}[\phi ]\ln (P_{n_{k}n_{k%
{\acute{}}%
}....}^{\mathrm{out}}[\phi ]/\left\vert \Psi _{n_{k}n_{k%
{\acute{}}%
}....}[\phi ]\right\vert ^{2})
\end{equation*}%
For the whole ensemble, the mean hidden-variable entropy is%
\begin{equation*}
S_{\mathrm{hv}}^{\mathrm{out}}=\sum_{n_{k}n_{k%
{\acute{}}%
}....}p(n_{k},n_{k%
{\acute{}}%
},....)S_{\mathrm{hv}}^{\mathrm{out,\ }n_{k}n_{k%
{\acute{}}%
}....}
\end{equation*}%
where%
\begin{equation*}
p(n_{k},n_{k%
{\acute{}}%
},....)=p(n_{k})p(n_{k%
{\acute{}}%
})....
\end{equation*}

If the initial state is pure and in quantum equilibrium (so that $S_{\mathrm{%
hv}}^{\mathrm{in}}=S_{\mathrm{vonN}}^{\mathrm{in}}=0$), imposing the
conservation rule (\ref{ConsS}) implies%
\begin{equation}
\sum_{n_{k}n_{k%
{\acute{}}%
}....}p(n_{k},n_{k%
{\acute{}}%
},....)S_{\mathrm{hv}}^{\mathrm{out,\ }n_{k}n_{k%
{\acute{}}%
}....}=-\sum_{k}\left( \frac{\omega }{T}(e^{\omega /T}-1)^{-1}-\ln
(1-e^{-\omega /T})\right)  \label{log2k}
\end{equation}%
(where the sums are taken over outgoing modes only). This gives a constraint
on the subensemble distributions $P_{n_{k}n_{k%
{\acute{}}%
}....}^{\mathrm{out}}[\phi ]$ and on the total (nonequilibrium) distribution%
\begin{equation*}
P^{\mathrm{out}}[\phi ]=\sum_{n_{k}n_{k%
{\acute{}}%
}....}p(n_{k},n_{k%
{\acute{}}%
},....)P_{n_{k}n_{k%
{\acute{}}%
}....}^{\mathrm{out}}[\phi ]
\end{equation*}

As discussed in section 3, it is expected that the generation of quantum
nonequilibrium is driven by some unknown phenomenon taking place inside the
hole (perhaps near the singularity), which is communicated to the exterior
-- at the nonlocal hidden-variable level -- via the entangled state.

Note that in pilot-wave field theory, the outcomes of quantum measurements
depend on the initial field configuration $\phi $; and the probabilities for
the outcomes depend on the probability distribution $P[\phi ]$ for $\phi $.
In general, a nonequilibrium distribution of fields will imply a non-quantum
distribution of outcomes of quantum measurements. (For some simple examples
of measurement in pilot-wave field theory, see ref. \cite{BHK}.) Thus, the
electromagnetic field in quantum nonequilibrium will yield anomalous
polarisation probabilities for single photons, as discussed in sections 2.2
and 3.

According to our argument, then, particles radiated by black holes will show
statistical anomalies outside the formalism of quantum theory. These
deviations from standard quantum probabilities may be interpreted as
information leaking nonlocally from behind the event horizon, as discussed
above.

\section{Relating Hidden-Variable and von Neumann Entropies}

We have proposed that the pure-to-mixed transition envisaged by Hawking is
accompanied by the generation of quantum nonequilibrium. And, as a simple
quantitative hypothesis, we have assumed that the associated decrease in
hidden-variable entropy $S_{\mathrm{hv}}$ balances the increase in von
Neumann entropy $S_{\mathrm{vonN}}$. As we have noted, this simple
hypothesis involves a comparison of two very different forms of entropy.
Here we outline how, even in non-gravitational physics, there are
relationships (whose details remain to be developed) between $S_{\mathrm{hv}%
} $ and $S_{\mathrm{vonN}}$ -- the two kinds of entropy are not completely
independent.

Consider a statistical mixture of pure quantum states $\left\vert \psi
_{1}\right\rangle $ and $\left\vert \psi _{2}\right\rangle $ for some
system, with density operator%
\begin{equation*}
\hat{\rho}_{\mathrm{sys}}=\frac{1}{2}\left\vert \psi _{1}\right\rangle
\langle \psi _{1}|+\frac{1}{2}\left\vert \psi _{2}\right\rangle \langle \psi
_{2}|
\end{equation*}%
If the states are orthogonal, $\langle \psi _{1}|\psi _{2}\rangle =0$, then
according to quantum theory it is possible to separate the mixed ensemble
into pure subensembles with density operators $\hat{\rho}_{1}=\left\vert
\psi _{1}\right\rangle \langle \psi _{1}|$ and $\hat{\rho}_{2}=\left\vert
\psi _{2}\right\rangle \langle \psi _{2}|$. For example, an appropriate
coupling to an apparatus in an initial pure state $\left\vert
g_{0}\right\rangle $ leads to the evolution%
\begin{equation*}
\hat{\rho}_{\mathrm{total}}^{\mathrm{in}}=\hat{\rho}_{\mathrm{sys}}\otimes
\left\vert g_{0}\right\rangle \langle g_{0}|\rightarrow \hat{\rho}_{\mathrm{%
total}}^{\mathrm{out}}=\frac{1}{2}\left\vert \psi _{1}g_{1}\right\rangle
\langle \psi _{1}g_{1}|+\frac{1}{2}\left\vert \psi _{2}g_{2}\right\rangle
\langle \psi _{2}g_{2}|
\end{equation*}%
where $\langle g_{1}|g_{2}\rangle =0$. Conditionalising on the final
readings of the apparatus, we are left with two pure subensembles $\hat{\rho}%
_{1}$, $\hat{\rho}_{2}$ labelled by those readings. (The von Neumann entropy
for the whole ensemble is preserved in this process, being equal to $\ln 2$
throughout; while the sum of the reduced von Neumann entropies for the
system and apparatus -- associated with their reduced density operators --
increases from $\ln 2$ to $2\ln 2$, as a result of ignoring the correlations
generated by the interaction.)

If on the other hand $\langle \psi _{1}|\psi _{2}\rangle \neq 0$, no quantum
process can separate the subensembles, since no unitary (norm-preserving)
evolution can lead to both $\left\vert \psi _{1}g_{0}\right\rangle
\rightarrow \left\vert \psi _{1}g_{1}\right\rangle $ and $\left\vert \psi
_{2}g_{0}\right\rangle \rightarrow \left\vert \psi _{2}g_{2}\right\rangle $
with $\langle g_{1}|g_{2}\rangle =0$. As is well-known, in quantum theory it
is not possible to distinguish non-orthogonal states without disturbing them 
\cite{NC}.

Such a process \textit{can} exist, however, in nonequilibrium de
Broglie-Bohm theory (for example). Given a nonequilibrium ensemble of
systems whose statistical dispersion is less-than-quantum, these systems may
be used to perform `subquantum measurements' on ordinary (equilibrium)
quantum systems, allowing non-orthogonal quantum states to be resolved
without disturbing them \cite{Pram}. The key point is that de Broglie-Bohm
trajectories generally differ for non-orthogonal states, and in
nonequilibrium we can have more information about the trajectories than
quantum theory allows.

Let us illustrate this with a simple example. Take the system to be a free
(nonrelativistic) particle in one dimension with coordinate $x$, and with
alternative initial states $\left\vert \psi _{1}\right\rangle =\left\vert
p\right\rangle $, $\left\vert \psi _{2}\right\rangle =$ $\frac{1}{\sqrt{2}}%
\left( \left\vert p\right\rangle +\left\vert -p\right\rangle \right) $ or
wave functions%
\begin{equation*}
\psi _{1}(x,0)=\frac{1}{\sqrt{L}}e^{ipx},\;\;\;\psi _{2}(x,0)=\frac{1}{\sqrt{%
2L}}\left( e^{ipx}+e^{-ipx}\right)
\end{equation*}%
(where $L$ is a normalisation length). The apparatus pointer has coordinate $%
y$ and initial wave function%
\begin{equation}
g_{0}(y)=(2\pi \Delta ^{2})^{-1/4}e^{-y^{2}/4\Delta ^{2}}  \label{Gauss}
\end{equation}%
We assume that over an ensemble $y$ has an initial (presumed known)
nonequilibrium distribution $\pi _{0}(y)\neq \left\vert g_{0}(y)\right\vert
^{2}$.\footnote{%
We are assuming, for theoretical purposes, that the nonequilibrium
distribution is known. How such an ensemble could be discovered in practice
is a separate issue. One possibility, already mentioned, is that
nonequilibrium relic particles might be left over from the early universe 
\cite{AVIsch} -- where the nonequilibrium distribution of the parent
population could be deduced from measurements made on a random sample \cite%
{Pram}.} Further, for simplicity we consider the extreme case of an
essentially dispersion-free distribution $\pi _{0}(y)$, for which the values
of $y$ are concentrated arbitrarily closely to $y_{0}=0$. We shall show
that, for an appropriate interaction between $x$ and $y$, after an
arbitrarily short time a mixture%
\begin{equation}
\hat{\rho}_{\mathrm{total}}^{\mathrm{in}}=\left( \frac{1}{2}\left\vert \psi
_{1}\right\rangle \langle \psi _{1}|+\frac{1}{2}\left\vert \psi
_{2}\right\rangle \langle \psi _{2}|\right) \otimes \left\vert
g_{0}\right\rangle \langle g_{0}|  \label{Mixin}
\end{equation}%
(with $\langle \psi _{1}|\psi _{2}\rangle \neq 0$) separates into pure
subensembles -- with density operators $\hat{\rho}_{1}=\left\vert \psi
_{1}\right\rangle \langle \psi _{1}|$ and $\hat{\rho}_{2}=\left\vert \psi
_{2}\right\rangle \langle \psi _{2}|$ -- labelled by distinct values of the
pointer coordinate $y$. This separation generates an effective decrease in
von Neumann entropy, from an initial positive value to a final value of $0$.

To see this, at $t=0$ switch on an interaction $\hat{H}=a\hat{p}_{x}\hat{p}%
_{y}$ between $x$ and $y$. As in section 3, $a$ is a coupling constant and
we neglect the Hamiltonians of $x$ and $y$. An initial pure state $\Psi
_{0}(x,y)=\psi _{0}(x)g_{0}(y)$ evolves into $\Psi (x,y,t)$ according to the
Schr\"{o}dinger equation%
\begin{equation*}
i\frac{\partial \Psi }{\partial t}=-a\frac{\partial ^{2}\Psi }{\partial
x\partial y}
\end{equation*}%
The associated continuity equation%
\begin{equation*}
\frac{\partial \left\vert \Psi \right\vert ^{2}}{\partial t}+\frac{\partial 
}{\partial x}\left( \left\vert \Psi \right\vert ^{2}a\frac{\partial S}{%
\partial y}\right) +\frac{\partial }{\partial y}\left( \left\vert \Psi
\right\vert ^{2}a\frac{\partial S}{\partial x}\right) =0
\end{equation*}%
implies the de Broglie guidance equations\footnote{%
Note that for this Hamiltonian the roles of $x$ and $y$ are reversed with
respect to the gradient of phase.}%
\begin{equation*}
\dot{x}=a\frac{\partial S}{\partial y},\;\;\;\;\dot{y}=a\frac{\partial S}{%
\partial x}
\end{equation*}%
for the hidden-variable trajectories. For $\psi _{0}(x)=\psi _{1}(x,0)$ we
have%
\begin{equation*}
\Psi _{1}(x,y,t)=\frac{1}{\sqrt{L}}e^{ipx}g_{0}(y-apt)
\end{equation*}%
while for $\psi _{0}(x)=\psi _{2}(x,0)$ we have%
\begin{equation*}
\Psi _{2}(x,y,t)=\frac{1}{\sqrt{2L}}e^{ipx}g_{0}(y-apt)+\frac{1}{\sqrt{2L}}%
e^{-ipx}g_{0}(y+apt)
\end{equation*}%
To leading order in $at$, the phase $S=\func{Im}\ln \Psi $ is either%
\begin{equation*}
S_{1}=px
\end{equation*}%
or%
\begin{equation*}
S_{2}=\frac{atpy}{2\Delta ^{2}}\tan px
\end{equation*}%
The solution for the trajectory $y(t)$ (with initial condition $y_{0}=0$) is
then either%
\begin{equation*}
y_{1}(t)=apt
\end{equation*}%
or%
\begin{equation*}
y_{2}(t)=0
\end{equation*}%
After an arbitrarily short time the quantum density operator $\hat{\rho}_{%
\mathrm{total}}$ (following the usual unitary evolution) will still be given
by (\ref{Mixin}), so that $\hat{\rho}_{\mathrm{total}}^{\mathrm{out}}=\hat{%
\rho}_{\mathrm{total}}^{\mathrm{in}}$, to arbitrary accuracy. And yet, no
matter how small $t$ may be, if $\pi _{0}(y)$ is sufficiently peaked around $%
y=0$, the ensemble will have divided into subensembles labelled by distinct
pointer positions $y_{1}=apt$ and $y_{2}=0$, with system wave functions $%
\psi _{1}$ and $\psi _{2}$ respectively. (The key point here is that, for
small $at$, while the total quantum state of system and apparatus is hardly
affected, for sufficiently narrow $\pi _{0}(y)$ the small change in $y(t)$
at the hidden-variable level can be large enough to provide unambiguous
information about the initial quantum state of the system.)

The initial quantum state (\ref{Mixin}) has a total von Neumann entropy%
\begin{equation}
S_{\mathrm{vonN}}^{\mathrm{in}}=-\lambda _{+}\ln \lambda _{+}-\lambda
_{-}\ln \lambda _{-}  \label{vonNMixin}
\end{equation}%
where $\lambda _{\pm }=\frac{1}{2}\pm \frac{1}{2}|\langle \psi _{1}|\psi
_{2}\rangle |$ are the eigenvalues of $\hat{\rho}_{\mathrm{total}}^{\mathrm{%
in}}$. Mathematically, the quantity $-Tr(\hat{\rho}_{\mathrm{total}}\ln \hat{%
\rho}_{\mathrm{total}})$ is preserved by unitary evolution (even over finite
times), and at the end of the above (brief) interaction what one usually
calls the von Neumann entropy is strictly speaking still given by (\ref%
{vonNMixin}). Physically, however, the ensemble has separated into pure
subensembles identifiable by distinct pointer positions, so that in effect
the total von Neumann entropy has really become $S_{\mathrm{vonN}}^{\mathrm{%
out}}=0$.

This finite decrease in an arbitrarily short time is possible if the initial
apparatus ensemble has an arbitrarily narrow nonequilibrium distribution $%
\pi _{0}(y)$, corresponding to a hidden-variable entropy $S_{\mathrm{hv}}^{%
\mathrm{in}}$ of arbitrarily large magnitude. For example, with the Gaussian
wave function (\ref{Gauss}) of width $\Delta $, a Gaussian distribution $\pi
_{0}(y)$ of width $w$ has a relative entropy%
\begin{equation*}
S_{\mathrm{hv}}^{\mathrm{in}}=\frac{1}{2}\left( 1-w^{2}/\Delta ^{2}\right)
+\ln \left( w/\Delta \right)
\end{equation*}%
and $S_{\mathrm{hv}}^{\mathrm{in}}\rightarrow -\infty $ as $w\rightarrow 0$
(for fixed $\Delta $).

For the realistic case of finite $w<\Delta $, the initial hidden-variable
entropy $S_{\mathrm{hv}}^{\mathrm{in}}$ is finite, and the above separation
into pure subensembles could take place only to some approximation (in some
finite time $t$). The magnitude of $S_{\mathrm{hv}}^{\mathrm{in}}$ might be
regarded as the `cost' of (approximate) non-orthogonal state separation. But
the details of the efficiency of such separation -- for example, how large $%
\left\vert S_{\mathrm{hv}}^{\mathrm{in}}\right\vert $ must be to separate a
given mixture of non-orthogonal states with a given accuracy -- remain to be
developed.

What is required here is the development of a `thermodynamics' of quantum
nonequilibrium systems interacting with ordinary quantum equilibrium
systems. In a specific model such as de Broglie-Bohm theory, this should be
possible. By reasoning along the above lines, one may for example study the
interconversion of hidden-variable and von Neumann entropies, and elucidate
the relations between them. At present, such an extended `thermodynamics' is
in its infancy. But it is already clear that the presence of quantum
nonequilibrium, with $S_{\mathrm{hv}}\neq 0$, results in anomalous behaviour
for $S_{\mathrm{vonN}}$.

The von Neumann entropy $-Tr(\hat{\rho}\ln \hat{\rho})$ is usually
identified with thermodynamic entropy. However, von Neumann's original proof 
\cite{vN} really shows that $-Tr(\hat{\rho}\ln \hat{\rho})$ is the entropy
of a mixture of eigenstates of $\hat{\rho}$ with probabilities equal to the
corresponding eigenvalues.\footnote{%
The proof proceeds by separating the (orthogonal) components of the mixture
by means of semipermeable membranes.} It is then assumed that $-Tr(\hat{\rho}%
\ln \hat{\rho})$ is still the entropy for any mixture with the same density
operator $\hat{\rho}$; that is, it is assumed that the entropy depends only
on $\hat{\rho}$. This assumption is justified in quantum theory, where $\hat{%
\rho}$ provides complete statistical information about the ensemble. But as
we saw in section 2.3, in quantum nonequilibrium $\hat{\rho}$ fails to
provide complete statistical information. There is therefore no reason to
assume that ensembles with the same $\hat{\rho}$ will have the same physical
properties, and no reason to assume that the entropy depends on $\hat{\rho}$
alone. It therefore seems likely that the definition of thermodynamic
entropy will have to be generalised in quantum nonequilibrium.

A similar conclusion was drawn by Weinberg \cite{Wein} in the context of
nonlinear quantum theory, in response to Peres' claim \cite{Per89} that the
(hypothetical) nonlinear evolution of quantum states would spontaneously
decrease the entropy of a closed system. Note also that Peres \cite{Per}
(following von Neumann \cite{vN}) has considered a thermodynamic cycle,
involving a gas of photons with non-orthogonal polarisation states, and
hypothetical semipermeable membranes capable of separating the
non-orthogonal states. The apparent effect of the cycle is to convert heat
into work at a single temperature, in violation of the second law of
thermodynamics. As noted by Peres, the conclusion that the second law is
violated rests on the assumption that von Neumann entropy is equivalent to
thermodynamic entropy. But there is no reason why this should be true in a
hypothetical `postquantum' theory (such as nonequilibrium de Broglie-Bohm)
that allows such membranes to exist.

In the context of black holes and pure-to-mixed transitions, it is possible
that considerations of black-hole thermodyamics will lead to a natural
candidate for a generalised notion of thermodynamic entropy, which includes
contributions from hidden variables as well as from the usual quantum
degrees of freedom. Whether or not a generalised second law would imply the
production of quantum nonequilibrium by black holes remains to be seen.

In the context of deterministic hidden variables, it seems natural to assume
that any nonequilibrium distribution generated by a black hole should
contain the `lost' information regarding the initial quantum state. That is,
it seems natural to assume that there exists a retrodictive mapping from the
final nonequilibrium mixed state to the initial equilibrium pure state.
Imposing the existence of such a mapping may provide a guide to the
construction of an appropriate generalised entropy. This remains to be
studied.

In our discussion of black-hole evaporation, we made the simple assumption
that $S_{\mathrm{hv}}+S_{\mathrm{vonN}}$ is conserved. This assumption has
no firm theoretical basis at present, and may well prove to be wrong, even
if black holes really do generate quantum nonequilibrium. On the other hand,
it is already clear from non-gravitational physics that there are good
physical reasons for positing a connection between $S_{\mathrm{hv}}$ and $S_{%
\mathrm{vonN}}$. The two entropies are not as unrelated as they might appear
at first sight, making it not implausible that the evaporation of black
holes effectively converts hidden-variable entropy into von Neumann entropy,
as we have suggested. In any case, the assumption that $S_{\mathrm{hv}}+S_{%
\mathrm{vonN}}$ is conserved might be amenable to experimental test, as we
shall now discuss.

\section{Possible Experimental Tests}

As discussed in section 3, a characteristic feature of nonequilibrium
photons would be a breakdown of the quantum modulation (\ref{cos2}) of
transmission through a pair of polarisers. If the polarisers are set at a
relative angle $\Theta $, quantum theory predicts a transmission probability 
$\cos ^{2}\Theta $ through the second polariser, independently of the state
of the incoming photons. Generically in nonequilibrium, there will be
deviations from $\cos ^{2}\Theta $ as the angle $\Theta $ is varied \cite%
{Sig}.

Experiments were carried out by Papaliolios \cite{Pap}, with ordinary
laboratory photons, to test for deviations from $\cos ^{2}\Theta $
transmission for successive polarisation measurements over very short
timescales. Such deviations were predicted by Bohm and Bub \cite{BB}, in a
hidden-variable model where measurements generate quantum nonequilibrium for
short times. (The model is not of the type being considered in this paper,
where initial equilibrium remains in equilibrium in all non-gravitational
processes.) It was found that successive measurements within times of order $%
10^{-13}$ $\mathrm{s}$ led to agreement with the quantum $\cos ^{2}\Theta $
modulation to within $1\%$ \cite{Pap}.

The prediction that Hawking radiation will be in quantum nonequilibrium
might in principle be tested, if such radiation is ever detected from the
evaporation of microscopic black holes. A more promising experiment,
however, which might be feasible now, involves the prediction that if one
half of an entangled pair falls into a macroscopic black hole then quantum
nonequilibrium will be generated in the other half. This might be tested
using entangled photons generated by atomic cascade emission in black-hole
accretion discs.

It might be thought that nonequilibrium would in any case be smeared out by
the finite size of the emitting region. If photons are emitted with a
nonequilibrium distribution $\rho (\lambda )\neq \rho _{\mathrm{eq}}(\lambda
)$ that depends on the spatial location of the emission, then the ensemble
of received photons will have a distribution $\bar{\rho}(\lambda )$
consisting of a spatial average of $\rho (\lambda )$, and it might be that $%
\bar{\rho}(\lambda )$ is very close to $\rho _{\mathrm{eq}}(\lambda )$ even
if $\rho (\lambda )$ is not. While this might occur, it is not necessary.
And such averaging will have no effect at all if $\rho (\lambda )$ is
uncorrelated with spatial location. For example, in de Broglie-Bohm theory,
the outcome of a spin measurement for a nonrelativistic spin-1/2 particle is
determined by the initial position inside the two-component wave packet \cite%
{Bell}, and a nonequilibrium distribution will generally yield spin
probabilities deviating from quantum theory. If the position of each
particle is defined relative to its packet, any averaging over the positions
of the packets will not affect the deviation from quantum probabilities.
More generally, the hidden variables (whatever they might be) determining
the outcome of a photon polarisation measurement could be (at least
effectively) independent of where the photon is generated in space -- in
which case spatial averaging over the source will not affect any
nonequilibrium that may be present.

\subsection{Hawking Radiation from Microscopic Black Holes}

Black holes with a range of masses may have formed in the early universe 
\cite{ZN67,Hawk71}. The evaporation timescale (\ref{tscale}) is of order the
current age of the universe for microscopic black holes with $M\sim 10^{15}\ 
\mathrm{g}$ or $M\sim 10^{-18}M_{\odot }$. Thus, if primordial black holes
with mass $\lesssim 10^{15}\ \mathrm{g}$ are sufficiently abundant, it might
be possible to detect Hawking radiation from them today. The radiated power
scales with time as $\gtrsim (-t)^{-2/3}$, diverging at $t=0$, when the hole
is assumed to disappear in an explosion, whose products depend on the
high-temperature behaviour of matter.

For black holes of mass $M\sim 10^{15}\ \mathrm{g}$, a significant fraction
of the luminosity is expected to be in the form of $\gamma $-rays peaked at $%
\sim 100\ \mathrm{MeV}$ \cite{Page}. However, from measurements of the
diffuse $\gamma $-ray background, the mean density of primordial black holes
with mass $\lesssim 10^{15}\ \mathrm{g}$ is constrained to be no more than a
fraction of order $10^{-8}$ of the critical density of the Universe, making
direct detection difficult \cite{PH76,MacCarr,EGRET,CarrMac}. However, if
the black holes are sufficiently clustered in our own Galaxy, they might be
directly detectable \cite{Heck}. It has in fact been claimed that the
observed anisotropic component of the $\gamma $-ray background is caused by
primordial black holes clustered inside our Galactic halo \cite{W96}.
Further, it has been suggested that observations of a class of very
short-time $\gamma $-ray bursts are consistent with an interpretation as
primordial black holes evaporating in our Galaxy \cite{Cline}; though the
consistency of this scenario with the primordial density fluctuation
spectrum has been disputed \cite{Green}.

It remains to be seen if present or future $\gamma $-ray satellites will
lead to a definitive detection of radiation from primordial black holes. The
INTEGRAL mission is currently providing a wealth of new data, as in the
recent resolution of compact sources responsible for the soft $\gamma $-ray
glow from the Milky Way \cite{Lebrun}. Three new missions are imminent
(Astro E II, GLAST and Swift).

Should $\gamma $-rays from the evaporation of primordial black holes ever be
definitively detected, we suggest that their polarisation probabilities be
probed for deviations from the standard $\cos ^{2}\Theta $ modulation.
Polarimetry for $\gamma $-rays is currently at an early stage of
development, however. The polarisation of $\gamma $-rays may be measured by
Compton scattering. According to the theoretical differential cross-section
(the Klein-Nishina formula), linearly-polarised $\gamma $-rays Compton
scatter preferentially in the plane perpendicular to their axis of
polarisation. This technique has recently been applied to measure the linear
polarisation of the $\gamma $-ray burst GRB021206 (at energies $\sim 1$ $%
\mathrm{MeV}$) \cite{Coburn}, and is being developed further for future $%
\gamma $-ray telescopes which will be able to measure polarisation at
energies up to $20$ $\mathrm{MeV}$ \cite{Taj03}. Note that our experiment
requires two successive polarisation measurements (or a preparation followed
by a measurement) at a relative angle $\Theta $. Because Compton polarimetry
does not destroy but merely scatters the incoming photon, there seems no
reason in principle why this could not be done: if the (absorbing) photon
detector is removed from the first device, the second may be configured so
as to accept only a particular polarisation output ($\gamma $-rays scattered
in a particular direction) from the first.

In theories with large extra dimensions \cite{AHDD}, the Planck scale is of
order $1$ $\mathrm{TeV}$, raising the possibility that microscopic black
holes might be produced in collisions at the $\mathrm{TeV}$ scale. It is
expected that such holes will evaporate, or thermally decay, primarily into
standard-model particles (including hard photons) \cite{EHM}. Such events
could be observed at the Large Hadron Collider \cite{DL}, or in collisions
between cosmic rays and atmospheric nucleons \cite{FS,AG}. Anomalous
Centauro-like events in cosmic rays have in fact been interpreted in terms
of exploding microscopic black holes of mass $\sim 1$ $\mathrm{TeV}$ \cite%
{MMT}. Again, if particles from the Hawking decay of microscopic black holes
were definitively identified, we would suggest that their polarisation
probabilities be tested for deviations from quantum behaviour.

\subsection{Entangled Photons from Black-Hole Accretion Discs}

While microscopic black holes have not been definitively detected, the
existence of macroscopic black holes is well established. It is believed
that most (if not all) galactic nuclei contain a supermassive black hole, of
mass in the range $\sim 10^{6}-10^{10}\ $solar masses, and that our Galaxy
is populated with stellar-mass black holes. (For a review of the evidence,
see for example ref. \cite{CMS99}.) It is expected that most astrophysical
black holes are accompanied by thin accretion discs. The strong gravity
region close to the hole is characterised by the production of X-rays, and
the profile of an X-ray emission line from the inner region of the disc may
be used to probe the spacetime structure at the location of the radiating
material \cite{Fab89}.

Observation of the active galaxy MCG--6-30-15 by the X-ray satellite ASCA
detected a broadened and skewed K-shell X-ray (fluorescent) emission line of
iron, with a profile showing an extended red wing consistent with the effect
of gravitational redshift close to the horizon \cite{Tan95,Fab95}. The line
is believed to come from the inner region of the (thin) accretion disc, the
surface of which is irradiated by a continuum of X-rays (originating in a
hot corona above the disc), leading to photoionisation of iron. The
transition with the largest cross-section results in the ejection of a
K-shell ($n=1$) electron. An L-shell ($n=2$) electron can then drop into the
K-shell with the emission of a K$\alpha $ line (fluorescent) photon at $6.4$ 
$\mathrm{keV}$. The line is intrinsically narrow, but photons emitted at
different distances from the horizon suffer different gravitational
redshifts, resulting in a broad and skewed profile. Detailed calculations
(including relativistic Doppler effects) result in a line profile that
agrees strikingly with observation. Similarly broadened iron emission lines
have been detected from other (Seyfert) galaxies, as well as from black
holes in our Galaxy. (For reviews of fluorescent iron lines as probes of
black-hole systems, see refs. \cite{F00,RN03}.) Broadened lines from oxygen,
nitrogen and carbon have also been reported \cite{BR01,Ogle}; though this
interpretation has been disputed \cite{Lee01,Sako03}.

Now, a two-photon cascade emission (an atomic decay through an intermediate
state) generates a pair of photons with entangled polarisations. This effect
was used in the classic early tests of Bell's inequality, based on cascades
in atomic calcium and mercury \cite{CHSH69,FC72,C76,A82}. If such a cascade
could be identified sufficiently close to the horizon in a black-hole
accretion disc, then this naturally-occurring situation might be used to
realise the thought experiment considered in section 3. For there will be a
significant probability that one of the photons is captured while the other
is detected on Earth (or on an orbiting satellite). The polarisation
probabilities of the detected photons could then be tested for deviations
from the standard $\cos ^{2}\Theta $ modulation. In practice, however, a
number of difficulties must be addressed.

\begin{center}
\textit{Cascade Emission Close to the Event Horizon}
\end{center}

We require a two-photon cascade so close to the horizon that there is a
significant probability that one half of the entangled pair is actually
captured. It is helpful to first consider just how close to the horizon the
observed K$\alpha $ iron line originates from.

The iron line profile from MCG--6-30-15 was initially thought to be
consistent with a Schwarzchild black hole, with emission from the surface of
an accretion disc extending inwards to about $r=6M$, with the disc inclined
at about $30^{\circ }$ to the line of sight \cite{Tan95,Fab95}. However,
subsequent analysis and observation favour a near-extremal Kerr black hole ($%
a/M>0.94$, where $a\equiv J/M$ is the specific angular momentum), with line
emission from radii as small as $r\approx 2M$ \cite%
{Dab97,Y98,Wilms,Fab02,Rey04,VF04} -- very close to the horizon at $r=M+%
\sqrt{M^{2}-a^{2}}\lesssim 1.3M$. (Similar results have been obtained for
Galactic black holes \cite{M04a,M04b}.) Note that it is usually assumed that
the accretion disc does not extend all the way to the event horizon, but has
an inner edge close to or at the `radius of marginal stability' $r_{\mathrm{%
ms}}$ (the radius of the last stable circular orbit for massive test
particles). And in most models, the observed X-ray line emission cannot come
from inside $r=r_{\mathrm{ms}}$ \cite{RN03}.

If the interpretation of these observations is correct, the extreme red wing
of the K$\alpha $ iron line originates from radii within a factor of $2$ of
the horizon. For these photons, any entangled partners (should such exist)
will have a large probability of being captured by the black hole -- because
in a cascade the directions of the emitted photon momenta are not strongly
correlated (owing to atomic recoil), so that any partners could have been
emitted over a wide range of angles. (For photons coming from infinity the
capture cross section for a black hole of mass $M$ is $\sim 20\pi M^{2}$ 
\cite{BHP}.)

As for identifying a cascade close to the horizon, let us consider further
the observed K$\alpha $ iron line. The detected photons are produced by an
L-shell ($n=2$) electron dropping into the vacant K-shell ($n=1$), leaving a
vacancy in the L-shell, amounting to a vacancy transition $1s\rightarrow 2p$%
. A further radiative transition can then follow: for example, an M-shell ($%
n=3$) electron can fall into the L-shell, emitting a second (L$\alpha $)
photon, amounting to a vacancy transition $2p\rightarrow 3d$. According to
the detailed calculations of Jacobs \textit{et al}. \cite{Jac86}, for an
initial K-shell ($1s$) vacancy created in neutral iron, the probabilities
for the radiative vacancy transitions $1s\rightarrow 2p$ and $2p\rightarrow
3d$ are respectively $P(1s\rightarrow 2p)=0.28$ and $P(2p\rightarrow
3d)=0.18\times 10^{-2}$. (See table I of ref. \cite{Jac86}. The only other
possible transition after $1s\rightarrow 2p$ is $2p\rightarrow 3s$, whose
probability of $0.13\times 10^{-3}$ may be ignored here.) The probability
for the vacancy cascade $1s\rightarrow 2p\rightarrow 3d$ is then just $%
P(1s\rightarrow 2p\rightarrow 3d)=P(2p\rightarrow 3d)\approx 2\times 10^{-3}$%
, as $2p\rightarrow 3d$ can occur only after $1s\rightarrow 2p$. Thus, for
an initial ensemble of iron atoms with a K-shell ($1s$) vacancy, about $%
0.2\% $ will undergo $1s\rightarrow 2p\rightarrow 3d$, emitting a pair of
photons. Further, of the subensemble that emits a K$\alpha $ photon, a
fraction%
\begin{equation*}
P(2p\rightarrow 3d|1s\rightarrow 2p)=P(2p\rightarrow 3d)/P(1s\rightarrow
2p)\approx 6\times 10^{-3}
\end{equation*}%
will subsequently emit an accompanying L$\alpha $ photon. In other words, of
the K$\alpha $ photons currently being observed from black-hole accretion
discs, about $0.6\%$ are accompanied by L$\alpha $ cascade photons.\footnote{%
We have quoted the transition probabilities for the case of neutral iron
only. Results for other ionisation states are listed in ref. \cite{Jac86}.
The assumption that iron is neutral is thought to be a good approximation
for a wide range of accretion discs in active galactic nuclei, where the
temperature of the (surface of the) disc is expected to be low ($kT\sim 10$ $%
\mathrm{eV}$) -- in contrast with galactic black holes whose discs are
expected to have much higher temperatures ($kT\sim 1$ $\mathrm{keV}$) \cite%
{RN03}.} This figure is small, but significant.

Thus, there appears to be no difficulty in identifying cascade photons
emitted sufficiently close to the horizon. Given that relativistically
broadened lines are beginning to be reported from other elements besides
iron, the situation is likely to improve in the near future. (Broadened line
emission from oxygen in NGC 4051 has been reported to require emission down
to $r<1.7M$, for a near-extremal Kerr black hole \cite{Ogle}.)

In principle, the proposed experiment might be attempted now with the
observed iron K$\alpha $ photons. But even if the effect we are looking for
exists, and even if the cascade partners are highly entangled (which we have
not established), the effect will be greatly diluted because most ($99.4\%$)
of the K$\alpha $ photons are not accompanied by cascade partners and are
therefore not susceptible to the suggested deviations from quantum
probabilities. Further, there is a difficulty with performing the required
polarisation measurements for X-ray photons. An efficient X-ray polarimeter
(based on the direction of electron emission in the photoelectric effect)
has been developed for astrophysical observations in the $2-10$ $\mathrm{keV}
$ band \cite{Costa01}, and might be deployed in the proposed XEUS mission to
study supermassive black holes \cite{Par00}. This device could certainly be
used to measure the polarisation of the K$\alpha $ iron line (at $6.4$ $%
\mathrm{keV}$). However, our experiment requires two successive measurements
at a relative angle $\Theta $, and to achieve this an alternative method for
measuring X-ray polarisation would have to be developed -- for unlike the $%
\gamma $-ray Compton polarimeter mentioned above, a polarimeter based on the
photoelectric effect has the unfortunate feature of destroying the measured
photon (which is absorbed by an atom, resulting in the ejection of a
photoelectron).

\begin{center}
\textit{Maximising Quantum Information Loss and Bell Inequality Violation}
\end{center}

The proposed conservation rule (\ref{ConsS}) suggests that the effect we are
looking for will be large when the von Neumann entropy generated by tracing
over the infalling photons is large -- that is, when the quantum information
lost behind the horizon is large. To ensure that this is the case raises
some practical difficulties, to which we now turn.

For a mixed two-photon state with nonzero (total) von Neumann entropy, the
relevant quantity is the difference between the reduced and total von
Neumann entropies. In the simple (if rather unrealistic) pure case, the
relevant quantity is just the reduced von Neumann entropy, which is also a
good measure of entanglement for pure states. As we noted in section 3, for
pure states the amount of nonequilibrium predicted by (\ref{ConsS}) grows
monotonically with the maximal violation of Bell's inequality, in accord
with our intuitive picture of nonlocal information flow from behind the
horizon. Previous work on correlation experiments designed to violate Bell's
inequality is then a useful guide.

The theory of two-photon polarisation correlations in atomic cascades was
worked out in great detail by Fry \cite{Fry73}. The results were applied to
select an appropriate atomic transition for use in tests of Bell's
inequality. In a typical experiment performed in the laboratory, photon
detectors are placed collinearly on each side of the source, and for ideal
polarisers the (normalised) coincidence rate is calculated to be%
\begin{equation*}
\frac{1}{4}\left( 1+F(\delta )\cos 2\phi \right) 
\end{equation*}%
Here, $\delta $ is the half-angle subtended by the detectors and $\phi $ is
the relative angle between the polariser axes; $F$ depends on the details of
the transition, and is a measure of the degree of correlation between the
polarisations of the photons. The correlations are large enough to violate
Bell's inequality if and only if%
\begin{equation}
|F(\delta )|\geq 2^{-1/2}  \label{Sqrt1}
\end{equation}%
Because $\left\vert F(\delta )\right\vert $ is a monotonic decreasing
function \cite{Fry73}, we require%
\begin{equation}
|F(0)|\geq 2^{-1/2}  \label{Sqrt2}
\end{equation}%
As already noted, because of atomic recoil the directions of the emitted
momenta are not strongly correlated. If $\delta $ is increased, so as to
accept non-antiparallel pairs of momenta, $\left\vert F(\delta )\right\vert $
decreases and the polarisation correlation is reduced.

Fry has calculated and tabulated values of $F(0)$ for many cascade
transitions \cite{Fry73}. With appropriate preparation of the populations of
the initial states, there are many cascades satisfying (\ref{Sqrt2}).
However, as shown by Fry, if the initial states are isotropically populated
(with respect to the atomic angular momentum), then of the many possible
cascades only 5 satisfy (\ref{Sqrt2}). These all have zero nuclear spin:
otherwise the hyperfine structure weakens the polarisation correlations to a
level that can be explained locally. Examples are the $0-1-0$ cascade of
atomic calcium and the $1-1-0$ cascade of atomic mercury used in the
experiments cited above.\footnote{%
An initial excited atomic state has total electron angular momentum $J_{i}$.
A transition $J_{i}-J-J_{f}$ consists of decay through an intermediate state 
$J$ to a final state $J_{f}$.} At first sight it might seem likely that in
an accretion disc the atomic states would be isotropically populated --
however, magnetic fields in the disc might generate a non-isotropy.

Note that (in ideal conditions) the cascade $0-1-0$ induces a perfect
polarisation correlation by virtue of conservation of angular momentum and
of parity -- provided the detectors are placed so that the accepted photons
have anti-parallel (`back-to-back') momenta. Schematically (for the pure
case, and ignoring the atom), the emitted photon pair may be represented by
an entangled state of the form%
\begin{equation*}
\sum_{\mathbf{k}r,\ \mathbf{k}%
{\acute{}}%
r%
{\acute{}}%
}c(\mathbf{k}r,\mathbf{k}%
{\acute{}}%
r%
{\acute{}}%
)|\mathbf{k}r\rangle \otimes |\mathbf{k}%
{\acute{}}%
r%
{\acute{}}%
\rangle
\end{equation*}%
where $\mathbf{k},\mathbf{k}%
{\acute{}}%
$ are the emitted momenta and $r,r%
{\acute{}}%
$ the polarisations. For the $0-1-0$ cascade, the coefficients $c(\mathbf{k}%
r,\mathbf{k}%
{\acute{}}%
r%
{\acute{}}%
)$ contain terms of the form $(\mathbf{\varepsilon }_{\mathbf{k}r}\cdot 
\mathbf{\varepsilon }_{\mathbf{k}%
{\acute{}}%
r%
{\acute{}}%
})$ and $(\mathbf{\varepsilon }_{\mathbf{k}r}\cdot \mathbf{k}%
{\acute{}}%
)(\mathbf{\varepsilon }_{\mathbf{k}%
{\acute{}}%
r%
{\acute{}}%
}\cdot \mathbf{k})$ \cite{Ball2} (where the linearly-independent
polarisation vectors $\mathbf{\varepsilon }_{\mathbf{k}r}$, $r=1,2$, are
orthogonal to the momenta $\mathbf{k}$). For small acceptance angles the
relevant $\mathbf{k},\mathbf{k}%
{\acute{}}%
$ are nearly antiparallel; $(\mathbf{\varepsilon }_{\mathbf{k}r}\cdot 
\mathbf{k}%
{\acute{}}%
)(\mathbf{\varepsilon }_{\mathbf{k}%
{\acute{}}%
r%
{\acute{}}%
}\cdot \mathbf{k})$ is then small, while the presence of $(\mathbf{%
\varepsilon }_{\mathbf{k}r}\cdot \mathbf{\varepsilon }_{\mathbf{k}%
{\acute{}}%
r%
{\acute{}}%
})$ constrains the polarisation state to take the form%
\begin{equation}
\frac{1}{\sqrt{2}}\left( |\mathbf{\hat{x}}\rangle \otimes |\mathbf{\hat{x}}%
\rangle +|\mathbf{\hat{y}}\rangle \otimes |\mathbf{\hat{y}}\rangle \right)
\label{polpsi}
\end{equation}%
where $|\mathbf{\hat{x}}\rangle $, $|\mathbf{\hat{y}}\rangle $ respectively
denote polarisation along the $x$-, $y$-axes (taking $\mathbf{k},\mathbf{k}%
{\acute{}}%
$ respectively along $+z$, $-z$). If the detectors accept non-antiparallel
pairs $\mathbf{k},\mathbf{k}%
{\acute{}}%
$, terms in $(\mathbf{\varepsilon }_{\mathbf{k}r}\cdot \mathbf{k}%
{\acute{}}%
)(\mathbf{\varepsilon }_{\mathbf{k}%
{\acute{}}%
r%
{\acute{}}%
}\cdot \mathbf{k})$ reduce the polarisation correlation.

The required restriction on the angle depends on how quickly the function $%
\left\vert F(\delta )\right\vert $ decreases. For a $0-1-0$ cascade, (\ref%
{Sqrt1}) is satisfied for $\delta \lesssim 70%
{{}^\circ}%
$ (assuming perfect polarisers), while for other cascades the required range
can be much smaller (for example for $\frac{1}{2}-\frac{3}{2}-\frac{3}{2}$
the range is $\delta \lesssim 15%
{{}^\circ}%
$), and the largest range is obtained for the case $1-1-0$, for which $%
\delta \lesssim 95%
{{}^\circ}%
$. (See the plots in Fig. 3 of ref. \cite{Fry73}.) In the experiments
performed by Aspect \textit{et al}. (with a $0-1-0$ cascade in calcium), $%
\delta =32%
{{}^\circ}%
$ and $F(\delta )=0.984$, showing that large angles need not significantly
decrease the correlation \cite{A02}. Clearly, at least for some cascades, a
large range of emission angles does not diminish the correlations below the
bound required to violate Bell's inequality. It suffices that the momenta $%
\mathbf{k},\mathbf{k}%
{\acute{}}%
$ be only \textit{approximately} oppositely-directed.

The above concerns an experiment with collinear detectors on each side of
the source. In the situation at hand, we have just one detector at one wing
of the entangled state. The discussion in section 3 suggests that our
proposed effect will be large in a situation in which Bell's inequality is
potentially violated by a large amount -- potentially, that is, if an
additional experimenter behind the horizon were actually to measure the
polarisations of the infalling photons. The question is whether such a
situation (without the additional experimenter) could arise in a black-hole
accretion disc.

The single detector (on Earth or a satellite) is so far away from the source
that the half-angle $\delta $ subtended is virtually zero. However, in
general the detected photons will have partners whose momenta are not
necessarily even approximately oppositely directed. (Here, by `oppositely
directed' we mean as defined in the local Lorentz rest frame of the emitting
atom.) Should one wish to restrict the experiment to photons with
approximately oppositely-directed momenta -- a case that is particularly
well understood and is known to provide strong polarisation correlations --
then this may be arranged if the line of sight from emission to Earth runs
close to the plane of the accretion disc (so that our view of the disc is
edge-on). Of the photons detected on Earth, some will have partners that
actually fell behind the horizon, and the momenta of the partners will be
approximately oppositely directed. The rest of the detected photons will
have partners that did not fall behind the horizon, and the momenta of the
partners need not be at all oppositely directed; these detected photons are
expected to have standard polarisation probabilities, and their presence
will merely dilute the sought-for effect. Note that X-ray flares \cite{Bag01}
and infrared flares \cite{Gen03} have been observed from the supermassive
black hole at the centre of our Galaxy -- and in view of our location in the
Galactic plane, our Galaxy would be the ideal system with which to arrange a
line of sight parallel to the accretion disc (assuming the disc to be
co-aligned with the Galactic plane).

It might be thought that the natural conditions in an accretion disc would
be too uncontrolled to produce the required entanglement. By comparison, in
the later Bell experiments with atomic cascades the atoms were excited by
lasers \cite{A82}. However, this was only for the sake of efficiency: in the
earlier Bell experiments, the atoms were excited by the (filtered) continuum
output from an arc lamp \cite{FC72} and by electron bombardment \cite{C76}.
Irradiation by a broad continuum of frequencies -- such as the continuum of
X-rays striking the surface of an accretion disc -- will excite some of the
atoms to appropriate states, leading to the cascade emission of entangled
photon pairs. Note also that, for a $0-1-0$ cascade and for approximately
oppositely-directed pairs, conservation of angular momentum and of parity
fixes the relative phase between the terms in the polarisation state (\ref%
{polpsi}), so that the entanglement is necessarily phase coherent. (A
mixture of pure entangled states with random relative phases would of course
be equivalent to a mixture with no entanglement.)

An edge-on view of the disc raises the question of the possible effect of
scattering along the line of sight. Entanglement is surprisingly robust
against scattering \cite{Alt02,Vel03}; though the momentum spread of the
scattered states (generated by for example elastic scattering in a random
medium) does diminish the polarisation entanglement \cite{Vel04,Ai04}. Of
greater concern is that significant scattering might destroy any quantum
nonequilibrium that may have been generated in the exterior photons: by the
time they reach Earth, scattering along the line of sight may have caused
their hidden variables $\lambda $ to relax back to the quantum distribution $%
\rho _{\mathrm{eq}}(\lambda )$. In the de Broglie-Bohm model, for example,
scattering terms in the wave function create perturbations in the velocity
field that can drive a system back to quantum equilibrium, if the terms are
sufficiently large \cite{AVbook}. However, whether or not such relaxation
occurs will depend on the details of the hidden-variables theory, as well as
on the extent of scattering. Scattering by dust in the plane of the galaxy
in question could be avoided by using a cascade that emits infrared photons
(or possibly X-rays).

Alternatively, one might consider a situation where the relevant photons do
not have approximately oppositely-directed momenta. This would be the case
if the accretion disc were viewed face-on. From the point of view of
observation this would be an advantage, because according to current models
the accretion discs in active galactic nuclei are surrounded by a co-aligned
dusty torus, so that a clear line of sight to the central engine is obtained
only face-on \cite{Anton93}. However, this experiment would require a
cascade emission in which approximately orthogonal photon momenta $\mathbf{k}%
,\mathbf{k}%
{\acute{}}%
$ nevertheless yield strong polarisation correlations. This possibility
needs to be studied further.

In practice, polarisation correlations produced by atomic cascades are
usually considered for photon pairs with oppositely-directed momenta, but in
principle the results of Fry \cite{Fry73} may be applied more generally, as
well as to general atomic populations. In an accretion disc, the atomic
level population will of course be determined by local conditions such as
the temperature and the presence of magnetic fields. Again, this is a matter
for future research.

Note that in curved spacetime parallel transport is required to define
relations between spin directions at a distance \cite{Bor00}. The choice of
measurement axes required for a maximal violation of Bell's inequality must
be adjusted appropriately, and by this means the standard correlations and
maximal violations may be obtained even beyond the event horizon \cite{Ter04}%
. The need to adjust the axes is not relevant here, however, where we are
considering measurements at one wing only; it is only the \textit{potential}
violation of Bell's inequality that is desirable (were appropriate
measurements at the second wing actually carried out), and the maximum
possible violation is the same as in flat spacetime.

Note also that our scenario requires that the infalling photons do not
interact strongly with the infalling material, otherwise such interactions
might destroy the entanglement with the outgoing photons.

Finally, note that in the above discussion we have conflated two issues: the
degree of quantum information lost behind the horizon, and the degree of
(potential) violation of Bell's inequality. These are directly related in a
simple way for pure states (as noted in section 3), whereas for mixed states
the relationship is currently obscure.

\begin{center}
\textit{Further Remarks}
\end{center}

In any real experiment, polarisation measurements will always show
deviations from $\cos ^{2}\Theta $ resulting from ordinary noise and
experimental errors. This could be distinguished from the sought-for effect
by switching the input back and forth from the astronomical source to a
similar source (for example of iron atoms irradiated by X-rays) prepared in
the laboratory. Further, our discussion suggests that, if the proposed
effect exists, it will have a distinctive signature: the deviations from $%
\cos ^{2}\Theta $ should be bigger for those photons that are closer to the
red end of the emission line, as these are emitted closer to the horizon and
are therefore more likely to have partners that were actually captured.

Recent developments in X-ray interferometry achieve an angular resolution
which will, in the foreseeable future, make it possible to image
supermassive black holes directly \cite{Cash00} (as envisioned in the
projected MAXIM mission \cite{MAX}). Though as we have noted, to perform two
successive polarisation measurements for X-ray photons requires a
measurement technique that does not destroy the measured photon. Further, if
the experiment were indeed performed at X-ray wavelengths, it would have to
take place above the Earth's atmosphere (which is opaque to X-rays). A
ground-based experiment might be possible in the infrared, to which our
atmosphere is transparent in certain wavelength windows.

One should beware that, in principle, quantum nonequilibrium (if it exists)
could invalidate some of the standard techniques used to identify emission
lines. For example, diffraction-grating spectroscopy assumes that single
photons obey standard quantum scattering laws. If the nonequilibrium photon
statistics are sufficiently anomalous, instead of bunching in the strong
central maximum of the relevant beam (corresponding to a particular spectral
order), the photons might be scattered primarily into the weak side lobes --
that is, in directions that do not correspond to their true frequency.

Throughout our discussion, we have been assuming that the proposed effect
alters only the statistical distribution of outcomes of quantum measurements
-- and not the allowed values of those outcomes. (That is, we allow the
ensemble distribution $\rho (\lambda )$ of hidden variables to be anomalous,
but the mapping from $\lambda $ to outcome for a single system is
unchanged.) If, for instance, the effect should generate an unexpected shift
in photon frequencies, then this too could in principle lead to confusion as
to what has been observed.

Clearly there are a number of experimental difficulties and uncertainties,
and it is possible that even if the effect exists we would not detect it. On
the other hand, a positive detection would be of great interest. The
observation of anomalous polarisation probabilities (deviations from $\cos
^{2}\Theta $ transmission through two successive polarisers) for photons
from a distant source would be remarkable, regardless of any uncertainties
about the origin or nature of the photons. If such an effect were observed,
it may well prove possible to find an alternative explanation, and further
experiments would be required to decide between the alternatives.

The identification of an appropriate cascade is a task for the future. An
experiment could in principle be performed now with the K$\alpha $ iron
line. But a practical experiment will probably need to await the detection
of other relativistically broadened emission lines, with a larger fraction
of photons accompanied by entangled partners, and in a frequency band more
convenient for accurate measurements of polarisation.

\section{Discussion and Conclusion}

The hypothesis advanced in this paper is that, during the formation and
evaporation of a black hole, the extra degrees of freedom associated with
hidden-variables theories become unfrozen, resulting in deviations from
standard (Born-rule) quantum probabilities. In particular, we have suggested
that deviations from quantum theory will occur for particles outside the
horizon that are entangled with particles inside the horizon. Specifically,
in the case of photons we have proposed a search for anomalous polarisation
probabilities, in the form of deviations from the $\cos ^{2}\Theta $
modulation of transmission through a pair of polarisers at a relative angle $%
\Theta $. We have suggested that the effect might be observable
experimentally in Hawking radiation from primordial black holes, and in
entangled photons from atomic cascade emission in black-hole accretion
discs. For the latter in particular, we have considered a number of
practical difficulties, none of which seem insurmountable.

As we have noted, the generation of quantum nonequilibrium by black holes
opens up the possibility that the initial state could be retrodicted, so
that no information is lost. The initial pure state that collapses to form a
black hole might be recoverable, in principle, from the final nonequilibrium
Hawking radiation. This seems natural in the fundamentally deterministic
theories we have been considering. However, to know whether the process
really is reversible would require the construction of a detailed theory.

Today, we see thermal nonequilibrium because in the early universe
gravitation amplified the primordial inhomogeneities in energy density,
leading to the formation of large-scale structure \cite{Pad}. Were it not
for this peculiarity of gravitation, our universe would be in a state of
global thermal equilibrium. In contrast, with respect to hidden variables we 
\textit{do} see global equilibrium today: all systems we have access to obey
the Born rule (at least to high accuracy). Presumably then, there is no
hidden-variable analogue of the gravitational amplification of fluctuations.
However, according to our arguments, there are circumstances in which
gravitation can drive an ensemble of systems out of quantum equilibrium --
namely, in the presence of black holes. If the proposed process does exist,
then in the early universe the formation and evaporation of primordial
(microscopic) black holes will ensure that quantum nonequilibrium is present
at early times, as has been suggested elsewhere \cite%
{AVth,AVbook,PLA1,AVIsch,AV96,NS}.

If hidden variables do exist, one expects the Bekenstein bound on the number
of distinct states of a spatially bounded system to be merely a feature of
quantum equilibrium (another example of the `information compression' noted
in section 2.3). Out of equilibrium, extra degrees of freedom will be
unleashed. In the case of de Broglie-Bohm theory, the extra (currently
hidden) parameters are continuous configurations, which can in principle
store an unlimited amount of information.

For 30 years Hawking radiation has stood as a clue to some underlying
connection between quantum theory, gravitation and statistical physics.
According to the arguments given in this paper, the true nature of the
connection has hitherto not come to light because a crucial ingredient was
missing: the connection involves not ordinary statistical physics, but the
statistical physics of nonlocal hidden variables. If there are additional
variables outside the domain of quantum theory, then there is an additional
entropy reservoir that has not been taken into account. As we have seen,
this offers a new perspective on Hawking information loss: far from being
lost, information about systems behind the horizon can leak out in the form
of anomalous distributions at the hidden-variable level.

Should it prove possible in practice to identify atomic cascade photons that
form one half of an entangled ensemble, the other half having fallen into a
black hole, then the theoretical considerations of this paper suggest that
it would be worthwhile to test the received photons for deviations from the
Born rule -- for example, by searching for anomalous polarisation
probabilities. Even leaving aside the motivations put forward here, such an
experiment would be worthwhile on general grounds, as a test of quantum
theory in new and extreme conditions.

\textbf{Acknowledgement.} This paper is dedicated to the memory of Dennis W.
Sciama.

\end{document}